\documentclass[reprint,nofootinbib,showpacs,showkeys]{revtex4-1}% Physical Review D
\usepackage{palatino,url}
\usepackage{amsmath}
\DeclareMathOperator{\sign}{sign}
\usepackage{graphicx}
\usepackage{multirow}
\usepackage{bm}
\usepackage{tensor}
\def\naive{na\"{\i}ve}

\makeatletter
\def\p@subsection{\thesection-}
\def\p@subsubsection{\thesection-\thesubsection-}
\makeatother

\begin{document}

\title{Physical geometry of the quasispherical Szekeres models} 
\date{\today}
\author{Robert G. Buckley}
\email{robert.buckley@utsa.edu} 
\author{Eric M. Schlegel}
\email{eric.schlegel@utsa.edu}
\affiliation{"Department of Physics and Astronomy, University of Texas at San Antonio, San Antonio, Texas 78249, USA}

%Exact solutions in GR, Cosmology

\begin{abstract}

The quasispherical Szekeres metric is an exact solution to Einstein's equations describing an inhomogeneous and anisotropic cosmology. Though its governing equations are well-known, there are subtle, often-overlooked details in how the model's functions relate to its physical layout, including the shapes and relative positions of structures. We present an illustrated overview of the quasispherical Szekeres models and show exactly how the model functions relate to the physical shape and distribution of matter. In particular, we describe a shell rotation effect that has not previously been fully understood. We show how this effect relates to other known properties, and lay out some mathematical tools useful for constructing models and picturing them accurately.

\end{abstract}

\maketitle

\section{Introduction}

As cosmological observations become ever more precise, it becomes insufficient to describe structures and their evolutions by approximate means, such as perturbation theory (which fails to capture nonlinear structure growth at late times) and N-body simulations (which do not typically fully incorporate general relativity). For this reason, interest is growing in inhomogeneous models based on exact solutions to Einstein's equations. Although these models have restrictions in the shapes of structures allowed, they are useful because they allow direct, exact calculations of observables such as lensing convergence, angular diameter distance, and redshift, free from approximations or ambiguities. 

The most extensively studied inhomogeneous exact solution is the Lema\^{i}tre-Tolman (LT) class of models \cite{Lemaitre,Tolman}.\footnote{Some refer to these as Lema\^{i}tre-Tolman-Bondi (LTB) models, though Bondi's contribution came fourteen years after the initial discovery \cite{Bondi}.} This describes a spherically symmetric configuration of pressureless dust matter, with an optional cosmological constant. It provides a highly idealized simulation of simple structures such as cosmic voids, allowing us to calculate their observational effects. Many studies have examined LT models in various configurations \cite{Alnes,Alnes2,Enqvist,BiswasOnion,GBH,GBH2,Zumalacarregui,Marra,Marra2,Marra3,BolejkoFocusing,Szybka,Lavinto,ClarksonPerturbations,KrasinskiStructureForm,Mustapha,Celerier2,Bull,BolejkoInverse,Moss,Foreman,YooCMB,Tokutake,Moffat,Wang,BolejkoVoidForm,HellabyLTstructure,HellabyShellCrossings,ZibinPerturbations,ZibinDecayingModes,Redlich,Valkenburg,KrasinskiMisunderstandings}.  

However, the symmetry of the LT metric severely constrains the types of structures that it can model. The real universe contains structures that are not so simple, consisting of complex arrangements of walls and voids. The asymmetry of these arrangements can qualitatively alter the evolution and observational effects in ways the LT models cannot capture. A more advanced metric capable of better representing such structures is the quasispherical subclass of the Szekeres metric \cite{Szekeres}. This metric is a generalization to the LT metric, breaking the symmetry and allowing for much greater variety.

Many studies have examined Szekeres models and employed them to investigate various cosmological observables (e.g., \cite{BolejkoEvol,BolejkoSC,BolejkoSCCMB,BolejkoVolAv,BolejkoCoarse,PeelSC,TroxelLensing,KoksbangRayTrace,KoksbangSC,Mishra,Villani,Barrow2,Buckley}), and others have explored their general mathematical and geometric properties (e.g., \cite{Berger,Bonnor,Bonnor2,Barrow,GeorgSymmetry,Walters,HellabyRotation,HellabyWormholes,HellabyPoS,HellabyGeometry,HellabyLimits,Apostolopoulos,deSouza}). The mathematics involved in the models are well-developed enough to run simulations, generate data, and make predictions. Describing and visually depicting the actual shapes of structures in such simulations, though, remains a challenge. 

The purpose of this paper is to clarify and illustrate the geometry of the Szekeres models, showing the connections between the model functions and the physical picture the models represent. Some of this is merely elaborating on already-known features, but we also describe a shell rotation effect that has not yet been fully understood. We will show how this fits in with other properties of the model. We will also provide some mathematical tools for building and adjusting models, as well as for handling calculations within these models. 

This paper is organized as follows. In section~\ref{sec:ModelDefs} we lay out the metric and evolution equations defining the Szekeres models. We also explain the role of each function and how they relate to simpler models. We describe the two coordinate systems most commonly used in Szekeres models in section~\ref{sec:coords}, and show how they relate to one another. In section~\ref{sec:restrictions} we briefly go over some of the physical restrictions on the model functions. In section~\ref{sec:Efunction} we go into greater detail explaining how the asymmetric shape and physical properties of the metric relate to the model-defining functions. In section~\ref{sec:symmetry} we present special cases which result in partial symmetry, laying out the mathematical conditions and discussing some of the benefits of such models. In section~\ref{sec:methods} we provide some useful equations, including geodesic equations and coordinate transformations, as well as outlines of numerical methods that can be used to visualize the model's spatial arrangement. We show an alternative picture in section~\ref{sec:embedding}, in which the model is seen as a hypersurface embedded in a higher-dimensional background manifold. Finally, we make our closing discussion in section~\ref{sec:discussion}.

\section{\label{sec:ModelDefs}Basic model definitions}

The Szekeres metric is a generalization of the LT metric, introduced in 1975 by P. Szekeres \cite{Szekeres}. Like the LT metric, it contains only a comoving, irrotational, pressureless dust, and optionally a cosmological constant. Its constant-$t$ hypersurfaces are conformally flat \cite{Berger}, and it has no gravitational radiation due to a lack of changing quadrupoles \cite{Bonnor}, but unlike LT, the Szekeres metric does have dipoles in its matter distribution. In general, it has no symmetry; there are no Killing vectors, except in special cases \cite{Bonnor2}.

Szekeres models are described by the metric
\begin{align}
\label{eq:Szekeresmetric}
\mathrm{d}s^2 &= -\mathrm{d}t^2 + \frac{(R' - R \, \frac{E'}{E})^2}{\epsilon-k}\mathrm{d}r^2 + \frac{R^2}{E^2}(\mathrm{d}p^2 + \mathrm{d}q^2).
\end{align}
Primes denote partial derivatives with respect to the radial coordinate $r$, and $\epsilon = +1$, $0$, or $-1$, corresponding to the quasispherical, quasiplanar, and quasipseudospherical subtypes respectively. (It is possible for $\epsilon$ to change with $r$, resulting in multiple geometries joined together in one model, as described in \cite{HellabyGeometry}.) The quasispherical subtype has attracted the most attention, as it is simplest to understand, includes LT as a special case (allowing for direct comparisons), and is best suited to describing localized structures. From here on, when we refer to the Szekeres models, we will mean the quasispherical subtype.

Like LT, the Szekeres spacetime consists of a series of spherical shells labeled by the coordinate $r$, and the functions $R = R(t,r)$ and $k = k(r)$ play the same roles as in LT---that is, $R$ represents the proper areal radii of the shells, and $k$ is related to the local spatial curvature. The coordinates on the shell, $p$ and $q$, relate to the more familiar $\theta$ and $\phi$ by a stereographic projection, as we will explain in section~\ref{sec:coords}; the shells are still perfectly spherical. The only real difference comes from the function $E = E(r,p,q)$, which describes the departure from LT---unlike LT, the shells are not concentric, nor is matter distributed evenly across a given shell. The function $E(r,p,q)$ is defined in terms of three arbitrary functions of $r$ as
\begin{align}
\label{eq:Efunction}
E(r,p,q) = \frac{[p-P(r)]^2 + [q-Q(r)]^2 + \epsilon S(r)^2}{2S(r)}.
\end{align}
(We will hereafter omit the $\epsilon$ factor, as we are focusing on the $\epsilon = +1$ case.) We will refer to the functions $P(r)$, $Q(r)$, and $S(r)$ as the ``dipole functions''. Together, they describe a dipolar asymmetry that can change from shell to shell. This dipole structure was first demonstrated by de Souza \cite{deSouza}. The details of this asymmetry are subtle, and will be discussed in detail in section~\ref{sec:Efunction}. For now, we note that when $S' = P' = Q' = 0$ (meaning $E' = 0$ as well), the metric reduces to the LT metric.

Einstein's equations applied to this metric reduce to two useful equations. The first describes the evolution of $R$ over time:
\begin{align}
\label{eq:evolution}
\dot{R}(t,r)^2 &= \frac{2M(r)}{R(t,r)} - k(r) + \frac{1}{3}\Lambda \, R(t,r)^2, 
\end{align}
where a dot indicates a partial derivative with respect to $t$. The function $M(r)$ arises as an integration constant, and it gives the total effective gravitational mass inside the shell. $\Lambda$ is the usual cosmological constant. This equation has exactly the same form as the evolution equation for LT models, meaning that the asymmetries are arranged in such a way that they do not affect the evolution of the shells. (Apostolopoulos shows how this arises mathematically, through a decoupling of the evolution equations from the spatial divergence and curl equations \cite{Apostolopoulos}.) 

This equation also closely parallels the first Friedmann equation. Each shell therefore evolves like a slice of a pressureless Friedmann-Lema\^{i}tre-Robertson-Walker (FLRW) universe, albeit a different FLRW universe for each shell. $R$ evolves in a hyperbolic fashion where $k < 0$, parabolic where $k = 0$, and elliptic where $k > 0$. The solutions to Eq.~(\ref{eq:evolution}) are given in appendix~\ref{app:solutions}

The second relation from Einstein's equations gives the mass density:
\begin{align}
\label{eq:density}
4\pi\frac{G}{c^4}\rho(t,r,p,q) = \frac{M'(r) - 3M(r)\frac{E'(r,p,q)}{E(r,p,q)}}{R(t,r)^2\left[R'(t,r) - R(t,r)\frac{E'(r,p,q)}{E(r,p,q)}\right]}.
\end{align}
From here on, we will use units in which the gravitational constant $G$ and speed of light $c$ both equal unity. Note that the function $M(r)$ is different from the three-dimensional integral of $\rho$ (even though, as we will soon see, $E'/E$ averages to 0 across any shell), since Eq.~(\ref{eq:density}) omits the curvature factor from the metric (\ref{eq:Szekeresmetric}). Nevertheless, it is $M(r)$ that guides the gravitational evolution (\ref{eq:evolution}). This is why we call $M(r)$ the ``effective gravitational mass'' instead of the ``total mass''.

\begin{figure}[tbp]
\begin{center}
\includegraphics[width=8.5cm]{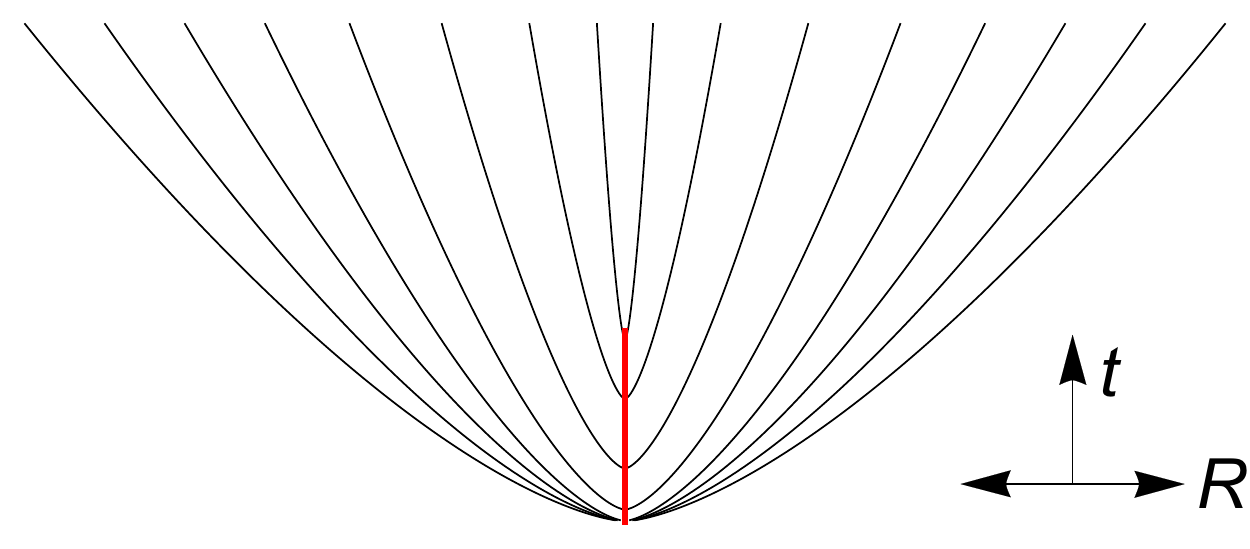} \\
\caption[Bang time]{An illustration of a non-simultaneous Big Bang, showing only one spatial dimension. Lines are surfaces of constant $r$. The thick red vertical line is the Big Bang singularity.}
\label{fig:bangtime}
\end{center}
\end{figure}

Integrating Eq.~(\ref{eq:evolution}) with respect to $t$ gives us another free function of $r$ as an integration constant:
\begin{align}
\label{eq:bangtime}
t - t_B(r) = \int_0^{R(t,r)}{\frac{\mathrm{d}\widetilde{R}}{\sqrt{2M(r)/\widetilde{R} - k(r) + \Lambda \, \widetilde{R}^2 /3}}}.
\end{align}
The function $t_B(r)$ is called the ``bang-time function,'' as it gives the time of the singularity $R = 0$ for any given shell.\footnote{It is possible to have collapsing solutions, in which case the RHS takes a negative sign and the function is better called the ``crunch-time function'', but this is simply a time reversal of the same situation.} Because the shells evolve independently from each other, they are mathematically allowed to begin their evolution at different times. That is, unlike the FLRW model which has all of space emerge from the Big Bang singularity simultaneously, Szekeres models (as well as LT models) can have different shells emerge at different times---outer shells can already be expanding while inner shells remain in the singularity, as shown in Fig.~\ref{fig:bangtime}.

We would appear to have six free functions of $r$: $M$, $k$, and $t_B$ shared with LT models, plus the three dipole functions. However, since $r$ is merely a label, it carries with it a gauge freedom; all of the above equations are invariant under transformations of the form $\widetilde{r} = f(r)$ (as long as $f$ is monotonic). This means that we can impose a definition of our choosing on one function of $r$, granting some welcome convenience. For instance, in some situations, it is advantageous to define $M(r) = r^3$, or even $M(r) = r$. Alternatively, we could define $R(t_0,r) = r$, where $t_0$ is the present time, which would make any one of the three LT free functions determined by knowledge of the other two by solving Eqs.~(\ref{eq:evolution}) and (\ref{eq:bangtime}). Any choice restricts the range of possible models somewhat---a choice of $M(r) = r^3$ cannot accommodate a model containing a vacuum over some range of $r$, and defining $R(t_0,r) = r$ rules out geometries in which $R$ is not monotonic in $r$, such as a closed universe or wormhole topology. But wide ranges of models remain available given any choice, and every model has gauge choices available. Regardless of our gauge choice, we end up with five functional degrees of freedom with which to define a particular model.

\section{\label{sec:coords}Coordinate systems} 

There are two different basic coordinate systems that are useful for handling the dimensions along the spherical surfaces: projective coordinates and spherical coordinates. Each has advantages that make it useful in different situations.

\subsection{Projective coordinates} 

\begin{figure}[tbp]
\begin{center}
\includegraphics[width=8.5cm]{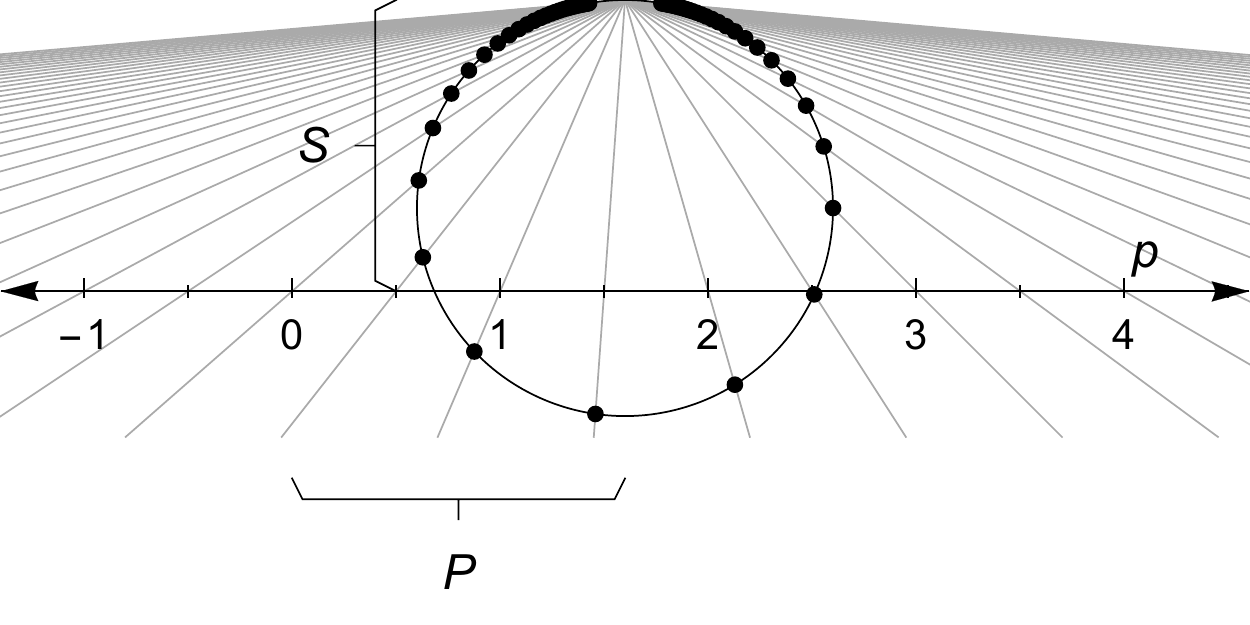} \\
\caption[Projective coordinates]{A 2D diagram showing how the projective coordinates map to the sphere. The dipole functions describe the negative offset of the projection plane from the projection point. The units are arbitrary, as long as the dipole functions and projective coordinates use the same units.}
\label{fig:projection}
\end{center}
\end{figure}

These are the most commonly used angular coordinates for Szekeres models, and the ones we have used in the previous section. Usually labeled $p$ and $q$, these coordinates map to the sphere by a simple Riemannian stereographic projection, as illustrated in Fig.~\ref{fig:projection}. Lines emerge from a projection point at the ``top'' the sphere, and intersect both the sphere and a two-dimensional (2D) projection plane at one point each. The plane is a distance $S(r)$ ``below'' the projection point (in arbitrary units), and it is marked by a Cartesian grid with its origin displaced by $(-P,-Q)$. Because these are functions of $r$, the coordinates can map differently on different shells.

At the projection point, the coordinates diverge. There is in general nothing physically special about this region, only a coordinate singularity, but this can cause problems for numerical calculations. And as we will see in section~\ref{sec:Efunction}, the location of this point can change from shell to shell; one shell's ``top'' does not necessarily point in the same direction as another's.

The convenience of these projective coordinates centers on the simplicity of the metric, Eq.~(\ref{eq:Szekeresmetric}). It is diagonal in this form, with reasonably simple components. It results in simple geodesic equations and Riemann and Ricci tensors (see appendix~\ref{app:curvature}), convenient for calculations. However, these coordinates are somewhat opaque to intuition. The shape of the $E$ function (\ref{eq:Efunction}) on the sphere, for instance, is not immediately obvious. We can tell that it has a minimum at $(p,q) = (P,Q)$, and increases away from this point (assuming $S$ is positive) proportionally to the square of the distance in the projective plane, but it is less obvious how this maps to the sphere.

More important than $E$ itself, though, is the combination $E'/E$. In projective coordinates, it takes the form
\begin{align}
\label{eq:EpEProjective}
\frac{E'}{E} = -2\frac{P'(p-P) + Q'(q-Q) - S'S}{(p-P)^2 + (q-Q)^2 + S^2} - \frac{S'}{S}.
\end{align}
The solution to $E'/E = 0$ traces a circle in the projective plane, centered at
\begin{align}
\label{eq:nullEpEcenter}
(p_c,q_c) &= \left(P-P'\frac{S}{S'}, Q-Q'\frac{S}{S'} \right),
\end{align}
with radius
\begin{align}
\label{eq:nullEpErad}
L &= \frac{S}{S'} \sqrt{P'^2 + Q'^2 + S'^2}.
\end{align}
In the special case $S' = 0$, the solution is a line instead of a circle. In either case, the line or circle demarcates a boundary between a region in which $E'/E$ is positive and one in which it is negative. See \cite{HellabyWormholes} for further discussion. 

\begin{figure}[tbp]
\begin{center}
(a)\\\includegraphics[width=8.5cm]{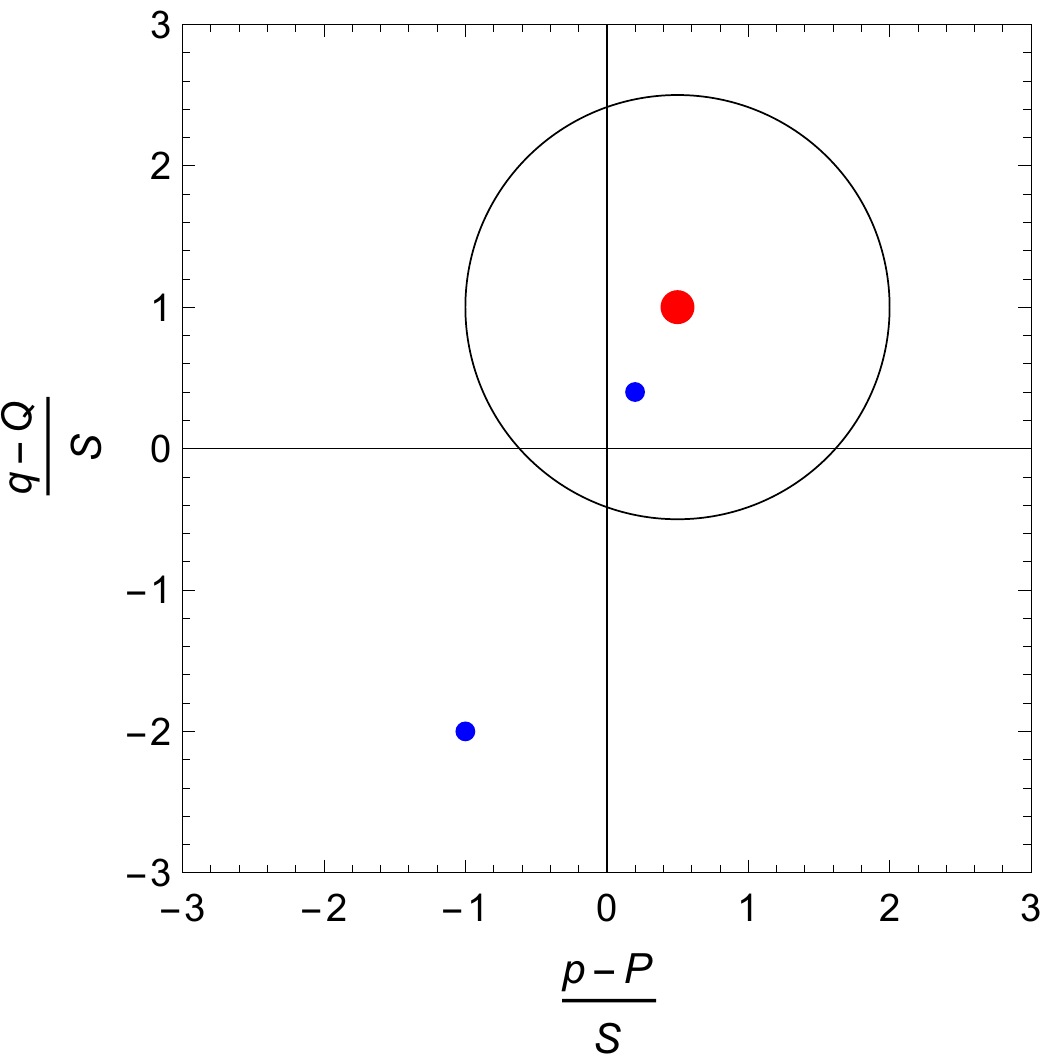} \hspace{1cm}
(b)\\\includegraphics[width=8.5cm]{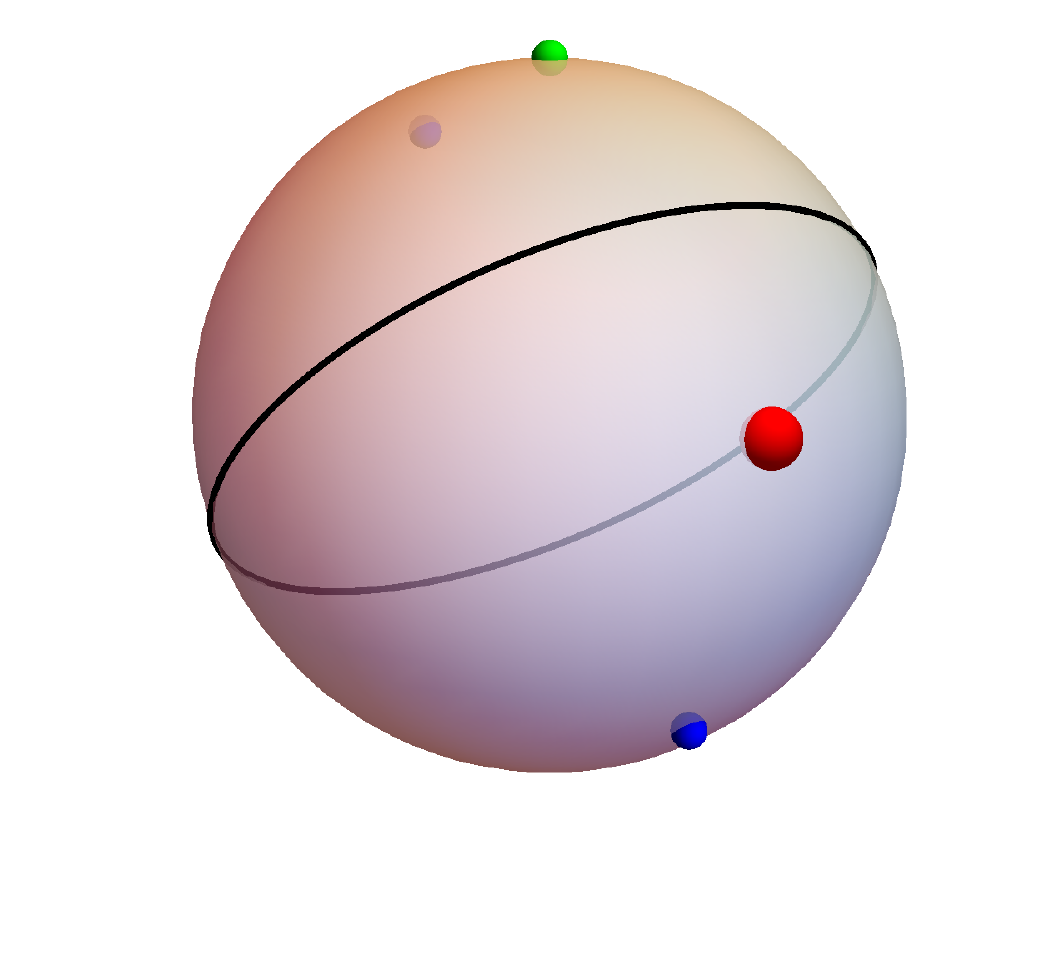}  
\caption[Great circle in the projective plane]{(a): How a great circle on the sphere appears in the projective plane. The large red dot is the center of the projective circle, whereas the two smaller blue dots are the two centers of the great circle on the sphere. (b): The same great circle on the sphere, with the same points marked. The green dot is the projection point.} 
\label{fig:greatcircle}
\end{center}
\end{figure}

A circle in a stereographic projection maps onto a circle on the sphere, and vice versa. The $E'/E = 0$ circle is special, though---it is a great circle on the sphere. It will be useful to note some properties of great circles and their mappings to the projective plane. A great circle in the projective plane is identified by a relation between the circle's center point and its radius: 
\begin{align}
\label{eq:greatcircle}
L_{\mathrm{gc}}^2 &= (p_c - P)^2 + (q_c - Q)^2 + S^2,
\end{align}
where $(p_c,q_c)$ is the center of the circle in the projective plane. This does not, however, correspond to the center of the circle on the sphere. On the sphere, there are two center points. The great circle can be seen as the intersection of the sphere with a plane through the sphere's center, and a line orthogonal to the plane through the sphere's center intersects the sphere at these two points, with projective coordinates
\begin{align}
\label{eq:greatcirclenormal}
(p_o - P, q_o - Q) &= \frac{-S^2 \pm S \; L_{\mathrm{gc}}}{(p_c - P)^2 + (q_c - Q)^2} (p_c - P, q_c - Q ).
\end{align}
We will soon see that for the $E'/E = 0$ great circle, these orthogonal points are the locations of the extrema of $E'/E$. The relative positions of these points in the projective plane are illustrated in Fig.~\ref{fig:greatcircle}.

It will also be useful to note that for any $(p_1,q_1)$, the antipodal point has coordinates
\begin{align}
\label{eq:antipodalpoints}
(p_2, q_2) &= (P,Q) - \frac{S^2(p_1 - P,q_1 - Q)}{(p_1 - P)^2 + (q_1 - Q)^2}.
\end{align}

\subsection{Spherical coordinates}

We can bring the coordinates to a more familiar form with a simple transformation:\footnote{A similar transformation using $\tan$ instead of $\cot$ is possible as well. This is equivalent, only with the projection point placed at $\theta = \pi$.}
\begin{subequations}
\label{eq:sphericaltransform}
\begin{align}
p - P &= S \cot\left(\frac{\theta}{2}\right) \cos \phi, \\
q - Q &= S \cot\left(\frac{\theta}{2}\right) \sin \phi.
\end{align}
\end{subequations}
The coordinates $\theta$ and $\phi$ describe the same geometry, but in terms of latitude and longitude instead of stereographic projection, with the pole $\theta = 0$ corresponding to the projection point, where $p$ and $q$ diverge. In these spherical polar coordinates, the metric is significantly more complicated and no longer diagonal:\footnote{The partial $r$ derivatives here are still taken with $p$ and $q$ held constant, {\itshape not} $\theta$ and $\phi$.} %
\begin{widetext}
\begin{align}
\label{eq:sphericalmetric}
\mathrm{d}s^2 &= -\mathrm{d}t^2 + \left[\frac{(R' - R\frac{E'}{E})^2}{1-k} + R^2 \left(1 - \cos \theta \right)^2 \left( \frac{P'^2 + Q'^2 + S'^2}{S^2} - \frac{2}{1 - \cos \theta} \frac{S'}{S} \frac{E'}{E} \right) \right] \mathrm{d}r^2 \nonumber \\
 &\ + 2 \frac{R^2 \sin \theta}{1 + \cos \theta} \left(\frac{E'}{E} - \frac{S'}{S} \right) \mathrm{d}r \, \mathrm{d}\theta + 2 \frac{R^2 \sin^3 \theta}{1 + \cos \theta} \left( \frac{Q'}{S} \cos \phi - \frac{P'}{S} \sin \phi \right) \mathrm{d}r \, \mathrm{d}\phi + R^2(\mathrm{d}\theta^2 + \sin^2 \theta \: \mathrm{d}\phi^2).
\end{align}
\end{widetext}
Though this appears far more complicated, in some ways these coordinates provide greater clarity. For instance, note that if we set $\mathrm{d}r = 0$, the spatial parts reduce to the metric of a 2-sphere, making it easier to see that surfaces of constant $r$ are indeed spherical (with radius $R$) regardless of anisotropies induced by $E$. The $E$ function itself takes a simpler form in spherical coordinates:
\begin{align}
\label{eq:ESpherical}
E(r,\theta,\phi) = \frac{S(r)}{1 - \cos \theta}.
\end{align}
While simpler, this form does not offer much immediate insight. Notice, though, that the metric in spherical coordinates only contains $E$ in the form of $E'/E$, as does the density equation (\ref{eq:density}). This important expression takes an evocative form in spherical coordinates:
\begin{align}
\label{eq:EpESpherical}
\frac{E'}{E} = -\frac{S' \cos \theta + (P' \cos \phi + Q' \sin \phi)\sin \theta}{S}.
\end{align}
This makes it clear that $P'/S$ defines an anisotropy in the direction of $(\theta, \phi) = (\pi/2, 0)$, $Q'/S$ in the direction $(\pi/2, \pi/2)$, and $S'/S$ in the direction $\theta = 0$---what we would call the ``$x$'', ``$y$'', and ``$z$'' directions in rectangular coordinates. While these rectangular coordinates are valid only on a given shell, not globally across the model, it will nevertheless sometimes be useful to refer to a ``{\itshape local rectangular frame}'' (or LRF).\footnote{We will refer to the LRF throughout this paper.} That is, in the LRF, 
\begin{align}
x \equiv \sin \theta \, \cos \phi, \qquad y \equiv \sin \theta \, \sin \phi, \qquad z \equiv \cos \theta. 
\end{align}

We can also see from Eq.~(\ref{eq:EpESpherical}) that $E'/E$ has a dipolar shape over the sphere (see \cite{Walters} for a detailed derivation), ranging from a maximum value of
\begin{align}
\label{eq:SzekExtreme}
\left(\frac{E'}{E} \right)_{\mathrm{max}} &= \frac{\sqrt{(P')^2 + (Q'^2) + (S')^2}}{S}
\end{align}
to the negative of this same value at the antipodal point. The directions of these angular extrema are given by		
\begin{subequations}
\label{eq:SzekExtremeDir}
\begin{align}
\theta_{\mathrm{max}} &= \cos^{-1}{\left(-\frac{S'}{\sqrt{P'^2+Q'^2+S'^2}}\right)}, \\
\phi_{\mathrm{max}} &= -\sign{Q'} \cos^{-1}{\left(-\frac{P'}{\sqrt{P'^2+Q'^2}}\right)}
\end{align}
\end{subequations}
(taking $S$ to be positive for all $r$). This is simply the set of spherical coordinates for the point $(x,y,z)_{\mathrm{max}} \linebreak = (P',Q',S')/\sqrt{P'^2 + Q'^2 + S'^2}$ in the LRF. 

The corresponding projective coordinates are 
\label{eq:SzekExtremeDirProj}
\begin{align}
(p,q)_{\mathrm{max}} &= (P,Q) - \frac{(P',Q')}{S'/S + \left(E'/E \right)_{\mathrm{max}}},
\end{align}
and $p_{\mathrm{min}}$ and $q_{\mathrm{min}}$ are related in the same way to $(E'/E)_{\mathrm{min}} = -(E'/E)_{\mathrm{max}}$. It is straightforward to see that these correspond to the great circle orthogonal points defined in Eq.~(\ref{eq:greatcirclenormal}) for the $E'/E = 0$ great circle, Eqs.~(\ref{eq:nullEpEcenter})--(\ref{eq:nullEpErad}).

We see that the geometry of the anisotropy on a single 2-sphere shell is quite simple. It is fully described by a dipole, symmetric about the two extrema and vanishing on the great circle midway between them. The magnitude and orientation of this dipole can change from one shell to the next, though, so the model as a whole is generally not symmetric.

\section{\label{sec:restrictions}Physical restrictions} 

While the Szekeres model has five functional degrees of freedom, the domains from which these functions can be chosen are not unlimited. They must satisfy certain conditions to avoid singularities and other pathological behavior. A few such restrictions are straightforward:
\begin{itemize}
	\item All of the free functions should be differentiable.
	\item In order to maintain the Lorentzian signature of the metric, we must have $k \leq 1$. Equality can only occur where $R' - R \: E'/E = 0$. Ensuring this holds for all $t$ and all $(p,q)$ requires  ${t_B' = P' = Q' = S' = 0}$ at this $r$. Furthermore, requiring finite density also means $M' = 0$, and continuity of $k'$ implies $k' = 0$. This set is a regular maximum or minimum (i.e. a belly or neck), which can occur in closed models or wormhole topologies. They are further discussed in \cite{HellabyWormholes}.
	\item Since shells have positive size, $R \geq 0$.
	\item $S \neq 0$, or else $E$ diverges and the projective mapping fails. Since $S$ must be continuous in consequence of being differentiable, this means that $S$ cannot change sign. By convention, we choose $S > 0$.
	\item There may be 0, 1, or 2 origins, meaning points at which $R$ vanishes besides the bang or crunch. At the origin(s), the density and curvature must be finite, and the evolution must smoothly match the surroundings. Hellaby and Krasi\'nski \cite{HellabyWormholes} have shown that regular origins require:
\begin{align}
\label{eq:originconditions}
M \sim R^3, \qquad k &\sim R^2, \qquad %\nonumber \\
(S,P,Q) \sim R^n,
\end{align}
where $n \geq 0$.
\end{itemize}
A few others require a little more explanation.

\subsection{Non-negative mass density}

Aside from a recent controversial proposal \cite{Farnes}, cosmology typically does not involve negative masses, and it is unclear how negative mass would behave gravitationally if it did exist. It is therefore common to require that the mass density is non-negative everywhere. In LT models, this simply means that $M'$ and $R'$ must have the same sign. In Szekeres models, this condition is still necessary, but no longer sufficient. The full condition is
\begin{align}
\label{eq:nonnegmass}
M' - 3M \, \frac{E'}{E} &\geq 0,
\end{align}
or $\leq 0$ if $R' - R E' / E \leq 0$ as well.

This can be translated into a direct restriction on the derivatives of the dipole functions. For Eq.~(\ref{eq:nonnegmass}) to hold across an entire shell, it suffices to satisfy the equation at the angular maximum of $E'/E$, which is given in Eq.~(\ref{eq:SzekExtreme}). We therefore have
\begin{align}
\label{eq:nonnegmass2}
\frac{\sqrt{(P')^2 + (Q'^2) + (S')^2}}{S} &\leq \left \vert \frac{M'}{3M} \right \vert.
\end{align}

\subsection{\label{sec:shellcrossing}Shell crossing}

In LT models, shell crossing may occur when $R' = 0$ at any $r$ and $t$. Unless this is a regular extremum, multiple shells are occupying the same space, causing the density to become infinite (unless $M' = 0$ there). This happens when one shell expands (or contracts) {\itshape through} another shell (and all shells in between). This would result in a coordinate degeneracy, with particles with different velocities at the same spatial location, violating the model's foundational assumption of comoving coordinates (as discussed in \cite{HellabyShellCrossings}). In order to maintain the model's validity, shell crossings are to be avoided.

In Szekeres models, shell crossing can occur even if $R'$ remains positive, if $R' - R \: E'/E = 0$ at any point in the full four-dimensional spacetime, besides at a regular extremum. Unlike in LT models, the singularity is not in general a full sphere, but rather begins as a point where the shells meet on one side. The surface of intersection grows from this point. On any given shell $r$, the shell crossing (if any) traces a circle parallel to the $E' = 0$ great circle.

We typically want to construct our models in such a way that no shell crossings occur, at least in the region of interest. The basic requirement is that there are no solutions to $R' \linebreak - R \: E'/E = 0$, which means
\begin{align}
\label{eq:shellcrossingprevention}
\frac{\sqrt{(P')^2 + (Q'^2) + (S')^2}}S &\leq \frac {R'}R.
\end{align}
For a more thorough discussion, including conditions on $M$, $k$, and $t_B$ which ensure Eq.~(\ref{eq:shellcrossingprevention}) holds globally, see \cite{HellabyWormholes}.

\section{\label{sec:Efunction}Effects of the dipole functions}

We have seen that the three dipole functions are solely responsible for the anisotropy in the model, each defining anisotropies in orthogonal directions. Although we have laid out their effects on the metric and on the density at a surface level, we have been vague so far about what is actually going on---what sort of physical anisotropies the dipole functions create. In this section we will build a more complete intuitive understanding of their physical effects, to the point where we can draw physically accurate pictures of the model.

The primary effects of the dipole functions are threefold: they shift the shells relative to each other, rotate their axes, and redistribute matter. We briefly explained these effects previously in \cite{Buckley}, but we shall examine them in greater depth here.

\subsection{\label{sec:shifting}Shell shifting}

\begin{figure}[tbp]
\begin{center}
\includegraphics[width=8.5cm]{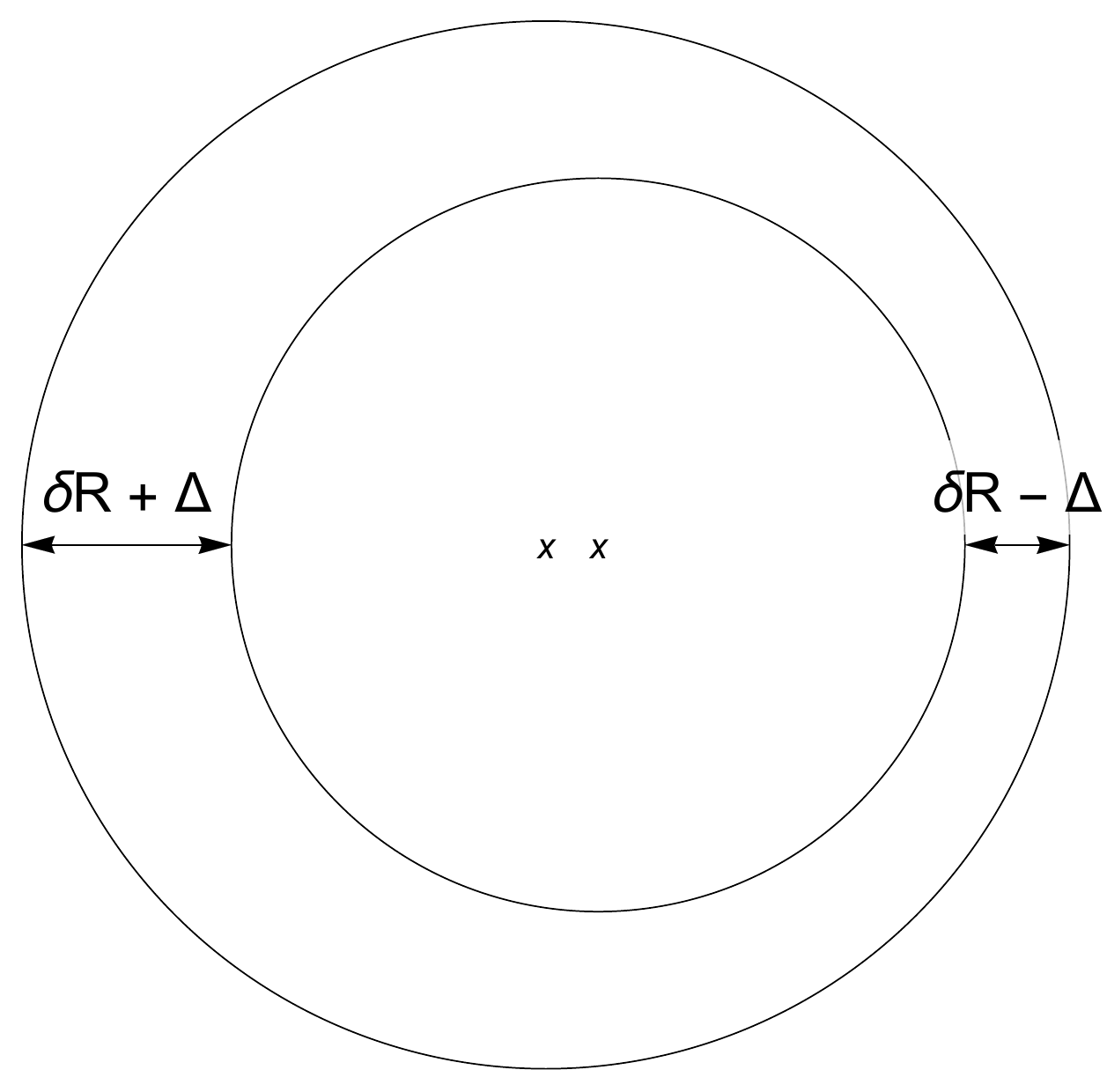} 
\caption[Shell shifting]{A simple illustration of how the dipole functions displace shells relative to each other. The geometric centers of the two shells are marked by 'x's. If the shifting between the two shells is all in the same direction, i.e. axially symmetric, the total shifting amount is $\Delta = \int{(R \: (E'/E)_{\mathrm{max}})/\sqrt{1-k} \; \mathrm{d}r}$ between the two shells.}
\label{fig:shellshifting}
\end{center}
\end{figure}

The first effect is that the dipole functions displace the shells relative to each other---that is, the shells are non-concentric, closer together on one side than the other. Due to the curved geometry, there is some ambiguity in how we measure the shifting distance, but the most straightforward measure is based on examining the $rr$ component of the metric in projective coordinates, Eq.~(\ref{eq:Szekeresmetric}). Because the metric is diagonal in these coordinates, the minimum proper distance along a line connecting an arbitrary point to a nearly-adjacent shell can directly be seen in $g_{rr}$. Without the dipole functions, this distance is simply $R' \: \mathrm{d}r/\sqrt{1-k}$. The dipole functions add only the term $-R \: E'/E/\sqrt{1-k}$. From the shape of $E'/E$ discussed in section~\ref{sec:coords}, we can see that this gives a dipolar modulation about $(\theta_{\mathrm{max}},\phi_{\mathrm{max}})$, defined in Eq.~(\ref{eq:SzekExtremeDir}). This corresponds to a relative displacement of the shells. From this viewpoint, the shell at $r + \delta r$ is shifted relative to the shell at $r$ in the direction $(\theta , \phi )=(\pi/2, 0)$ by a distance
\begin{align}
\label{eq:xshift}
\delta_x &= \frac{R \: P'/S}{\sqrt{1-k}}\delta r,
\end{align}
in the direction $(\pi/2, \pi/2)$ by 
\begin{align}
\label{eq:yshift}
\delta_y &= \frac{R \: Q'/S}{\sqrt{1-k}}\delta r,
\end{align}
and in the direction $(0, 0)$ by 
\begin{align}
\label{eq:zshift}
\delta_z &= \frac{R \: S'/S}{\sqrt{1-k}}\delta r,
\end{align}
where the ``$x$'', ``$y$'', and ``$z$'' subscripts refer to the axes of the shell's LRF. The shells are farther apart in the direction of the shifting, and closer together in the opposite direction. This is illustrated in Fig.~\ref{fig:shellshifting}.

When looking at how the shifting affects the transverse separation of points, the displacement appears to have a different magnitude, but the same direction. Specifically, it follows Eqs.~(\ref{eq:xshift})--(\ref{eq:zshift}) with the $1/\sqrt{1-k}$ factor omitted. We show how this contributes to the metric in spherical coordinates in appendix~\ref{app:rotationproof}.

The shifting effect can be considered to be responsible for the $E'/E$ term in the denominator of the mass density function, Eq.~(\ref{eq:density}). Where the shells are compressed, the density increases proportionally, and likewise it decreases where the shells are stretched apart. The term in the numerator is a separate effect, indicating that the matter distribution of each shell is not held fixed as they are shifted around. This is further discussed in section~\ref{sec:matterdistr}.

\begin{figure}[tbp]	
\begin{center}
\includegraphics[width=8.5cm]{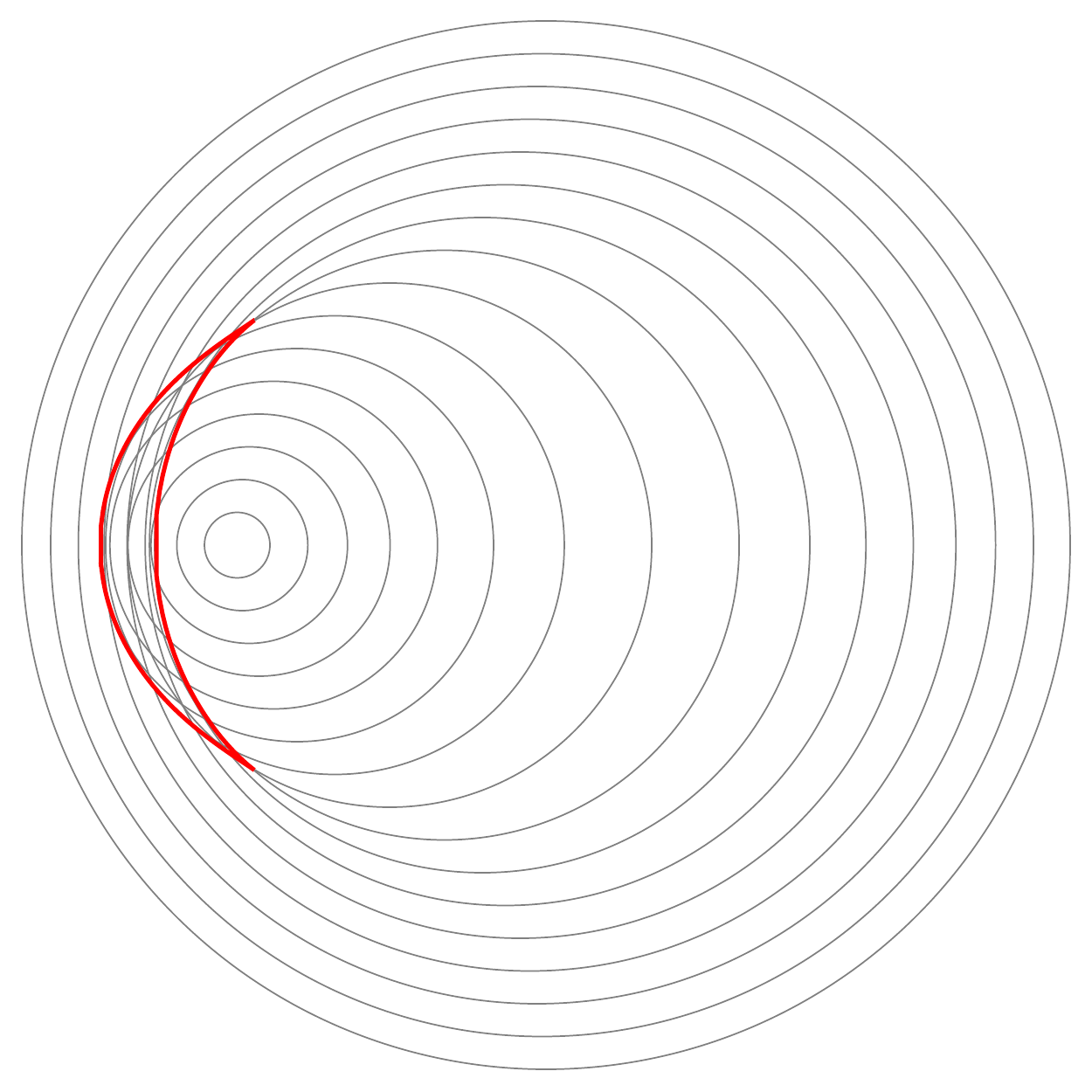} 
\caption[Shell crossing]{An example of shell crossing in a Szekeres model. The shell crossing surface is outlined in red. Inside this surface, the coordinates are degenerate.}
\label{fig:shellcrossing}
\end{center}
\end{figure}

The shell shifting effect makes it necessary to be especially careful of shell crossing singularities. We already saw in section~\ref{sec:shellcrossing} that Szekeres models run into shell crossings more easily than LT models. Now, we can more clearly picture why this is the case. Because the shells are non-concentric, smaller shells can intersect larger ones if we are not careful. The surface of intersection across all shells can form a non-trivial shape, an example of which is illustrated in Fig.~\ref{fig:shellcrossing}.

\subsection{\label{sec:rotation}Shell rotation}

The second effect is more subtle. It has been noted by Hellaby \cite{HellabyRotation} that the orientations of the coordinates appear to change from shell to shell, as evidenced by the rotation of the orthonormal tetrad along spatial paths. This property seems to have been overlooked by many authors. We have determined exactly how the shells' coordinates are rotated in terms of the dipole functions, which we briefly noted in \cite{Buckley}, and expand upon here.

The dipole functions change the shells' spherical coordinate frames as follows: the shell $r + \delta r$ is rotated relative to shell $r$ by $\delta r \: P'/S$ about the point $(\pi/2, -\pi/2)$ (the $-y$ axis in the shell's LRF) and by $\delta r \: Q'/S$ about the point $(\pi/2, 0)$ (the $x$ axis). A derivation showing that this rotation, combined with the shell shifting, results in the given metric in spherical coordinates, Eq.~(\ref{eq:sphericalmetric}), can be found in appendix~\ref{app:rotationproof}.

\begin{figure*}[tbp]
\begin{center}
(a) \hspace{5.1cm} (b) \hspace{5.1cm} (c) \hspace{0.4cm} ~ \\
\includegraphics[width=5.1cm]{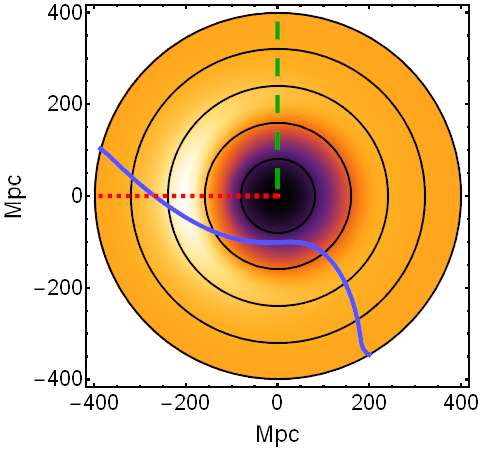} \hspace{0.4cm}
\includegraphics[width=5.1cm]{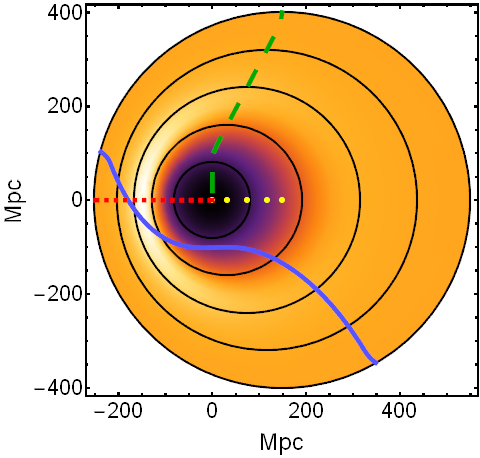} \hspace{0.4cm}  
\includegraphics[width=5.1cm]{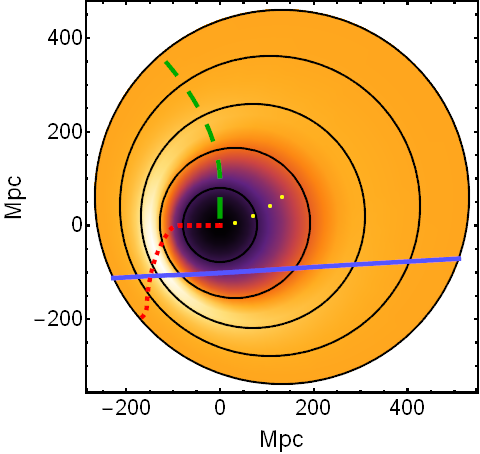}	\hspace{0.2cm}
\raisebox{0.6cm}{\includegraphics[width=1cm]{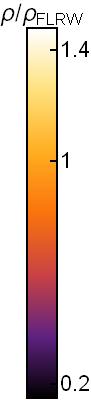}}
\caption[Effects of the dipole functions]{Three ways of plotting a cross-section of a void+wall Szekeres model, to illustrate the effects of the dipole functions on the coordinates. The only active dipole function is $P(r)$---the other two are constant. Detailed model definitions are given in appendix~\ref{app:example}. The blue line shows the path of an arbitrary null geodesic. Black circles mark shells of constant $r$ in steps of $80$ Mpc, and yellow dots mark the centers of these shells. The green dashed line marks $\theta = 0$ (the LRF's $+z$ axis at each shell), and the red dotted line marks $\theta_{\mathrm{max}}$, where the density has its angular maximum. (a): plotted in ``\naive'' coordinates, as though it was an LT model. Notice that the geodesic is not a straight line. (b): incorporating the shell shifting. The true width of the dense wall structure is more clearly seen, narrower than it seemed before. The geodesic still does not appear straight. (c): incorporating shell rotation. The wall is distorted, and the geodesic is now very nearly straight. This is the most physically accurate method of plotting in a two-dimensional image. The same method was used in the figures in the rest of this paper and \cite{Buckley}, and is explained in more detail in section~\ref{sec:plotting}.}
\label{fig:Szekeffects}
\end{center}

\end{figure*}

It may seem as though this is merely a coordinate effect, only present when using spherical coordinates. However, even in the projective coordinates, this effect is built into the behavior of the dipole functions. The directions along which the dipole functions act (the axes of the LRF) are rotated by this effect. For this reason, a model with only $S$ varying with $r$ ($P$ and $Q$ constant) is axially symmetric, whereas a model with only $P$ or $Q$ varying is not. For instance, if we set $S' = Q' = 0$, and $P' \neq 0$, anisotropies will all be oriented along the ``$x$'' direction in each shell's LRF, but the shell rotation will cause the ``$x$'' direction to change from shell to shell, resulting in the structures being smeared over a range of angles. Figure~\ref{fig:Szekeffects} illustrates such a case, showing how both the shifting and rotation affect the physical layout of the model. Note that with shifting alone, the geodesic line appears very curved, but it is nearly straight when the rotation is taken into account.\footnote{Although the geodesic's path covers a range of times, and the density plot shows a constant-time slice, the size of the structure is small enough that it does not evolve appreciably in the time it takes the geodesic to traverse it.}

\begin{figure}[tbp]
\begin{center}
\includegraphics[width=8.5cm]{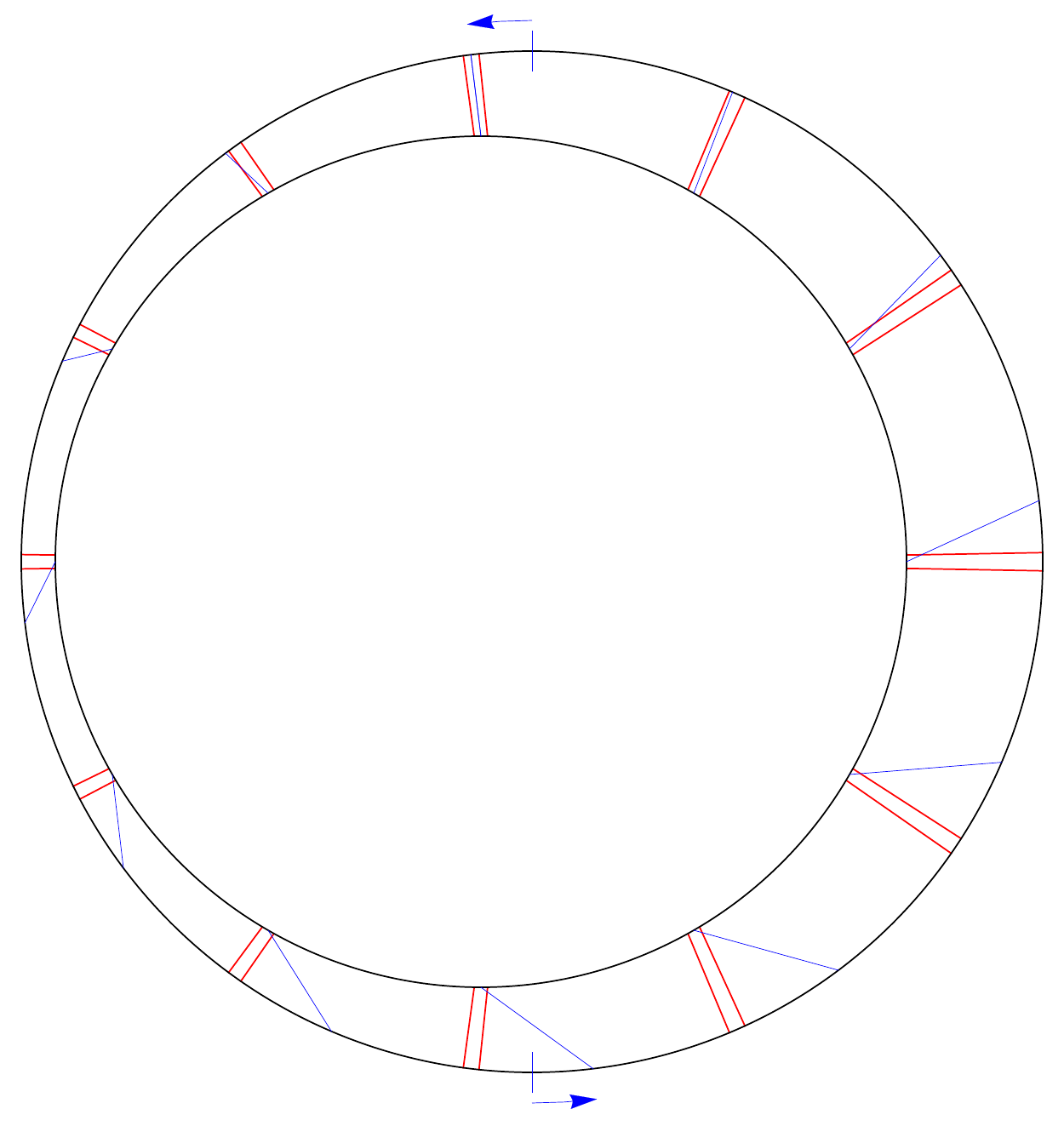} 	
\caption[Explanation of rotation effect]{A 2-dimensional illustration explaining how the rotation effect arises. Blue lines connect points of constant $\theta$. Double red lines connect points of constant $p$. Ticks on the outer shell mark where $\theta = 0$ and $\theta = \pi$ would be without the rotation effect, for comparison, and the arrows show the rotation.}
\label{fig:rotationexplanation}
\end{center}
\end{figure}

The reason this rotation effect is necessary can be explained in terms of the relation between the projective and spherical coordinates. Because the metric in projective coordinates, Eq.~(\ref{eq:Szekeresmetric}), is diagonal, we know that the shortest line connecting a point on a shell to a nearby shell is one with constant $p$ and $q$. This is not in general true for the spherical coordinates, with one important exception. The point $\theta = 0$ serves as the projection point, and therefore always corresponds to the point where $p$ and $q$ diverge. Indeed, since the off-diagonal terms $g_{r\theta}$ and $g_{r\phi}$ both include a $\sin \theta$ factor, the spherical metric is diagonal at this point. At $\theta = 0$, then, the line of constant $\theta$ and $\phi$ must also correspond to the shortest connecting line between two shells. Due to the shell shifting, this forces one to be rotated relative to the other, as illustrated in Fig.~\ref{fig:rotationexplanation}.

Note that the magnitudes of shifting, Eqs.~(\ref{eq:xshift})--(\ref{eq:zshift}) have a factor of $R(t,r)$, whereas the rotation magnitude does not. Unlike the shell shifting effect, the shell rotation is not time-dependent. The relative orientation of shells is maintained as they evolve. This missing factor also means that the shell rotation generally decreases in magnitude with increasing $r$, as $P'/S$ and $Q'/S$ are limited by the shell crossing condition, Eq.~(\ref{eq:shellcrossingprevention}).

\subsection{\label{sec:matterdistr}Matter distribution}

The third effect is not on the metric itself, but rather on the matter distribution. We saw how the shell shifting effect is reflected in the denominator of the density function, Eq.~(\ref{eq:density}), but this cannot explain the term in the numerator. This term represents a dipolar redistribution of matter on each shell, along the same direction as the shell shifting.

This redistribution is necessary to preserve the spherical shape and FLRW-like expansion of the shells. Even though the density in a shell's interior is not symmetric, it must always be arranged in a particular way such that the total gravitational effect on the shell allows it to expand uniformly, maintaining its spherical shape. This peculiar arrangement is what makes it possible for the model to be mathematically tractable.

The total impact on the density across a shell depends on how that shell's density in the corresponding LT model, $\rho_{\mathrm{LT}}$, compares to the effective average density in its interior, $\bar{\rho}_{\mathrm{int}}$. Specifically, if for a given shell
\begin{align}
\label{eq:densitycomparison}
\rho_{\mathrm{LT}} \equiv \frac{M'}{4\pi R^2 R'} &> \frac{3M}{4\pi R^3} \equiv \bar{\rho}_{\mathrm{int}},
\end{align}
then the overall density will be greatest on the side where the shells are compressed, and least on the opposite side. If, instead, the inequality is reversed, the density will be {\itshape less} where the shells are compressed and {\itshape greater} where they are stretched apart. The former case commonly occurs when modeling a void with secondary structures in and around it. Because a void's density profile increases from the center, every shell has a greater density than the interior. Therefore, Szekeres shifting results in a sharp, thin overdense wall on one side (narrow because of the shell compression), with a broader, shallower underdensity on the opposite side. (This is underdense relative to the rest of the shell, not relative to the interior.) Conversely, models with a central overdensity will see a perhaps less realistic arrangement of deep, narrow voids opposite thicker density bulges (again, relative to the shell, not the interior).

Suppose we hold the LT model functions unchanged in form while increasing the magnitudes of the dipole functions. As the shell shifting brings the shells close together on one side, the density contrast on the compressed side approaches positive or negative infinity, depending on whether the shell is overdense or underdense compared to the interior (or it stays constant, if $\rho_{\mathrm{LT}} = \bar{\rho}_{\mathrm{int}}$). Restricting the functions to being below the shell crossing limit, Eq.~(\ref{eq:shellcrossingprevention}), imposes a limit on the density on the stretched side, equal to halfway between $\rho_{\mathrm{LT}}$ and $\bar{\rho}_{\mathrm{int}}$:\footnote{In the case of an underdense shell, an even stronger limit is in place if we restrict the density to be everywhere non-negative, from Eq.~(\ref{eq:nonnegmass2}): $\rho_{\mathrm{max}} = 2 \rho_{\mathrm{LT}}/(1 + \rho_{\mathrm{LT}}/\bar{\rho}_{\mathrm{int}})$.}
\begin{align}
\rho_{\mathrm{min/max}} = \frac{M'}{8 \pi R^2 R'} + \frac{3M}{8\pi R^3}.
\end{align}
Whether this is an upper limit or a lower limit depends on how the shell's density compares to the interior. In either case, though, this means that a shell that is (over/under)dense anywhere is (over/under)dense everywhere, relative to the interior.%

If the two quantities $\rho_{\mathrm{LT}}$ and $\bar{\rho}_{\mathrm{int}}$ are equal, then the density on the shell will be uniform regardless of the dipole functions. This means that if we begin with a homogeneous background LT model (equivalent to FLRW), we cannot create inhomogeneities through the dipole functions alone. This presents a challenge if we wish to describe a complex arrangement of structures while maintaining homogeneity on large scales. To produce an overdensity on a section of a shell, the density across the {\itshape entire} shell must be greater than the average interior density. This can still be achieved without a single dominant central void (or overdensity), for instance by using an oscillating radial density profile, as in Sussman's prescription of periodic local homogeneity (PLH) \cite{Sussman}. 

In any case, the angular extrema of the density (and other scalar quantities) always coincide with the angular extrema of $E'/E$---the locations of maximum stretching and compression. Extending our view to the full three-dimensional (3D) space, the radial positions of the extrema are more difficult to find; $\rho' = 0$ is generally a difficult equation to solve, and some of the solutions are saddle points rather than maxima or minima (see \cite{Sussman} for further discussion). Furthermore, as this equation is time-dependent, the extrema are not generally comoving. Different scalars can have extrema at different positions, but all scalars have an extremum at the origin \cite{Sussman}.

\subsection{\label{sec:expansion}Secondary effect: expansion rates and shear}

\begin{figure}
\begin{center}
\includegraphics[width=8.5cm]{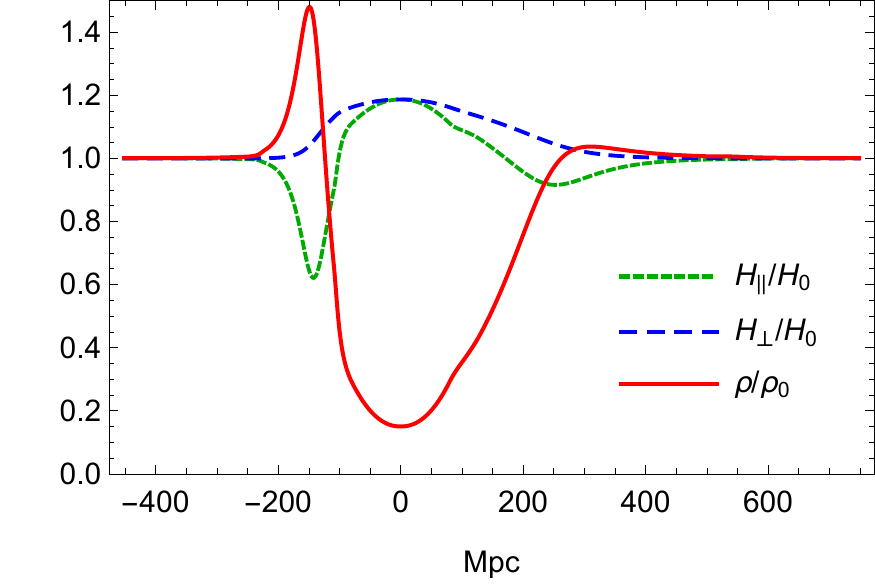} 	
\caption[Expansion rates]{Expansion rates and density profile in a one-dimensional slice through the center of a typical void-and-wall model, similar to that shown in Fig.~\ref{fig:Szekeffects}, except axially symmetric. Detailed model definitions are given in appendix~\ref{app:example}.}
\label{fig:expansionrates}
\end{center}
\end{figure}

In FLRW models, the expansion rate is characterized by a single Hubble value, the same in all locations and all directions.
\begin{align}
\theta_{\mathrm{FLRW}} = 3H_{\mathrm{FLRW}} = 3\frac{\dot{a}}{a}
\end{align}
In LT models, the expansion rate is modified not only by making it a function of the radial coordinate, but also by splitting it into two different rates: the transverse expansion $H_\perp$ (expansion in the two directions along the shell's surface) and the longitudinal expansion $H_\parallel$ (expansion between shells, along the line of sight of an observer at the center).
\begin{align}
H_{\perp,\mathrm{LT}} = \frac{\dot{R}}{R}, \\
H_{\parallel,\mathrm{LT}} = \frac{\dot{R}'}{R'}.
\end{align}
In Szekeres models, the transverse expansion rate is unchanged, because each shell evolves the same as in the corresponding LT model. The longitudinal (or radial) expansion, however, is modified due to the shell shifting.
\begin{align}
H_\parallel = \frac{\dot{R}' - \dot{R}\frac{E'}{E}}{R' - R \frac{E'}{E}}.
\end{align}
These two expansion rates are plotted in Fig.~\ref{fig:expansionrates} for an example void-and-wall model.

The overall expansion is given by the trace of the gradient of the matter velocity field, $\theta = u^\alpha{}_{;\alpha}$ (where a semicolon denotes a covariant derivative with respect to the following coordinate). In comoving coordinates (as we are using), the matter flow $u^\alpha$ simply equals $(1,0,0,0)$. In terms of $H_\parallel$ and $H_\perp$, the scalar expansion equals
\begin{align}
\theta = 2 H_\perp + H_\parallel.
\end{align}
This shows the overall rate of change of the size of a region. To see how the shape of a region changes, we look at the symmetric traceless component of the velocity field gradient, the shear, defined as $\sigma^\alpha{}_\beta = h^{\alpha \gamma}u_{(\gamma;\beta)} \linebreak - \frac{1}{3}\theta h^\alpha{}_\beta$, where $h^{\alpha \beta} = g^{\alpha \beta} + u^\alpha u^\beta$. In Szekeres models,
\begin{align}
\sigma^\alpha{}_\beta = \frac{1}{3}(H_\parallel - H_\perp)\mathrm{diag}(0,2,-1,-1).
\end{align}
The shear tells how space is stretched differently in different directions. The antisymmetric part of the velocity field gradient, the rotation, vanishes in Szekeres models:
\begin{align}
\omega^\alpha{}_\beta = h^{\alpha \gamma}u_{[\gamma;\beta]} = 0.
\end{align}
This is what is meant by the models being ``irrotational''. Because of this property, the Szekeres metric is incapable of modeling virialized structures. This is one of its most important limitations, but it allows the mathematics to remain tractable.

\section{\label{sec:symmetry}Symmetry conditions}

Most choices for the dipole functions result in models with no global symmetry at all. Setting all three to constants, however, reduces the metric to the LT metric, which possesses full spherical symmetry. There are also special cases with intermediate levels of symmetry.

\subsection{\label{sec:axialsymmetry}Axial symmetry}

If the Szekeres dipole lies along the same direction at all $r$, the overall model will be axially symmetric. Such a model will have a Killing vector field reflecting its single continuous rotational symmetry.

In some cases, it is advantageous to construct a model to be axially symmetric. While this reduces the freedom of the model, it also reduces the number of variables that must be considered, both in the model's definition and in simulations of observations. One can generate a complete picture of an axial observer's sky by only varying one angle in the geodesics. Furthermore, radial geodesics along an axis of symmetry have the special property of always staying on the axis, greatly simplifying the geodesic equation calculations, as we will show in section~\ref{sec:geodesics}. In fact, this is the only case in which a purely radial geodesic is possible in a Szekeres model \cite{Nolan}.\footnote{Further, Krasinski and Bolejko show that such radial geodesics are the only ones in Szekeres models that are ``repeatable light paths'', meaning two geodesics at different times can follow the same spatial path, resulting in no visible drift \cite{KrasinskiRLPs}.}

Because the $P$ and $Q$ functions rotate the frames of the shells relative to each other, a model with only $P'$ or only $Q'$ nonvanishing is not axially symmetric. A model with only $S'$ nonvanishing, however, is symmetric, with the axis of symmetry passing through $\theta = 0$ and $\theta = \pi$ on all shells. This is the simplest kind of axially symmetric model, having no shell rotation. Unfortunately, the $p$ and $q$ coordinates diverge on half of the symmetry axis, which can be inconvenient for calculations.

Models symmetric about other axes are possible, though not as simple. As shells are rotated relative to each other, the relative shifting direction must change to compensate, in order to maintain a straight line. Along a general symmetry axis, the spherical coordinates do not hold constant. The projective coordinates, though, {\itshape do} hold constant on the symmetry axis. This can be inferred from the diagonality of the metric in projective coordinates, Eq.~(\ref{eq:Szekeresmetric}). From any given point, the shortest distance to a nearby shell is along the path satisfying $\mathrm{d}p = \mathrm{d}q = 0$. If this point is on the symmetry axis, we can deduce from symmetry arguments that this minimal distance is along the axis; if it were in any other direction, the symmetry would be broken. We will use this property of the projective coordinates to derive conditions on the dipole functions which result in axial symmetry, similar to the presentation in \cite{SitUbEM}. Georg and Hellaby \cite{GeorgSymmetry} derive the same equations via the Killing equations.

The shifting direction on any shell corresponds to one of the two angular extrema of $E'/E$, which were given in Eq.~(\ref{eq:SzekExtremeDirProj}), but can also be found by solving 
\begin{align}
\label{eq:axialsymmetry}
E \, E_{,pr} &= E_{,p} E_{,r}, \qquad E \, E_{,qr} = E_{,q} E_{,r}.
\end{align}
Commas in subscripts denote partial derivatives with respect to the following coordinate(s). If these equations hold for the {\itshape same} $(p,q)$, which we will call $(p_0,q_0)$, across {\itshape all} $r$, the model is axially symmetric. (Such a model will also satisfy the equations for a second $(p,q)$, which we will call $(p_1,q_1)$, for the half of the symmetry axis on the opposite side of the origin.) 

Expanded in terms of the three dipole functions, these equations become
\begin{widetext}
\begin{subequations}
\label{eq:axialsymmetryExpanded}
\begin{align}
\label{eq:axialsymmetryExpandedP}
2(p_0 - P)S \, S' - 2(q_0 - Q)(p_0 - P)Q' &= \left[(p_0 - P)^2 - (q_0 - Q)^2 - S^2 \right] P', \\
\label{eq:axialsymmetryExpandedQ}
2(q_0 - Q)S \, S' - 2(q_0 - Q)(p_0 - P)P' &= \left[(q_0 - Q)^2 - (p_0 - P)^2 - S^2 \right] Q'.
\end{align}
\end{subequations}
\end{widetext}
As mentioned previously, these are immediately solved by taking $P' = Q' = 0$, with $P = p_0$ and $Q = q_0$. If $Q' = 0$ but $P' \neq 0$, the axial symmetry conditions reduce to
\begin{subequations}
\label{eq:axialsymmetryExpanded2}
\begin{align}
\label{eq:axialsymmetryExpanded2P}
2(p_0 - P)S \, S' = \left[(p_0 - P)^2 - (q_0 - Q)^2 - S^2 \right] P' , \\
\label{eq:axialsymmetryExpanded2Q}
(q_0 - Q)S \, S' - (q_0 - Q)(p_0 - P)P' = 0.
\end{align}
\end{subequations}
One way to satisfy Eq.~(\ref{eq:axialsymmetryExpanded2Q}) is to take $S \, S' = (p_0 - P)P'$. But if we then plug this into Eq.~(\ref{eq:axialsymmetryExpanded2P}), we get either $P'~=~0$ (and therefore $S'~=~0$) or
\begin{align}
\label{eq:axialsymmetryExpanded3P}
(p_0 - P)^2 + (q_0 - Q)^2 + S^2 = 0.
\end{align}
The former case is simply an LT model, whereas the latter can only be satisfied for real-valued functions if $S = 0$. This would result in rampant division-by-zero singularities in the metric and density function, so it is not a valid solution. We therefore go back to Eq.~(\ref{eq:axialsymmetryExpanded2Q}) and instead only consider the solution $Q = q_0$. We can then easily solve Eq.~(\ref{eq:axialsymmetryExpanded2P}) by relating the dipole functions by
\begin{align}
\label{eq:axialsymmetrysol1b}
S^2 &= C_2(p_0 - P) - (p_0 - P)^2,
\end{align}
where $C_2$ is an arbitrary constant. A similar solution is possible if $P' = 0$ but $Q' \neq 0$, by simply replacing $p_0$ and $P$ with $q_0$ and $Q$. 

If both $P'$ and $Q' \neq 0$, the axial symmetry conditions are satisfied by \cite{SitUbEM}
\begin{subequations}
\label{eq:axialsymmetrysol2}
\begin{align}
\label{eq:axialsymmetrysol2a}
q_0 - Q &= C_0(p_0 - P), \\
\label{eq:axialsymmetrysol2b}
S^2 &= C_1(p_0 - P) - (C_0^2 + 1)(p_0 - P)^2.
\end{align}
\end{subequations}
We can see that the case in Eq.~(\ref{eq:axialsymmetrysol1b}) is simply the special case when $C_0 = 0$. 

The symmetry axis passes through the origin, and $(p_0,q_0)$ covers only one side of it. The other side is antipodal to the first on each shell, so its coordinates can be found by applying Eq.~(\ref{eq:antipodalpoints}). In the case $P' = Q' = 0$, we mentioned that the projective coordinates diverge on the other side, but in the other cases, it has the coordinates
\begin{align}
p_1 &= p_0 - \frac{C_1}{C_0^2 + 1}, \qquad q_1 = q_0 - \frac{C_0 C_1}{C_0^2 + 1}.
\end{align}
For the special case of $P' = 0$, we need only set $C_0 = 0$ and then swap $p$ and $P$ with $q$ and $Q$.

The angular coordinates of the symmetry axis are determined by the constants $C_0$ and $C_1$ according to
\begin{subequations}
\label{eq:axialsymmetryangles}
\begin{align}
\label{eq:axialsymmetryanglesa}
C_0 &= \cot \phi_{\mathrm{ax}}, \\
\label{eq:axialsymmetryanglesb}
C_1 &= \frac{2S}{\sin \theta_{\mathrm{ax}} \cos \phi_{\mathrm{ax}}}.
\end{align}
\end{subequations}
We see that $\phi_{\mathrm{ax}}$ holds constant with $r$, while $\theta_{\mathrm{ax}}$ must vary as $S(r)$ varies. We also note that Eq.~(\ref{eq:axialsymmetryanglesa}) has two solutions, corresponding to the two sides of the symmetry axis, which also means Eq.~(\ref{eq:axialsymmetryanglesb}) has two corresponding solutions for $\theta_{\mathrm{ax}}$.

\begin{figure}[tbp]
\begin{center}
\includegraphics[width=8.5cm]{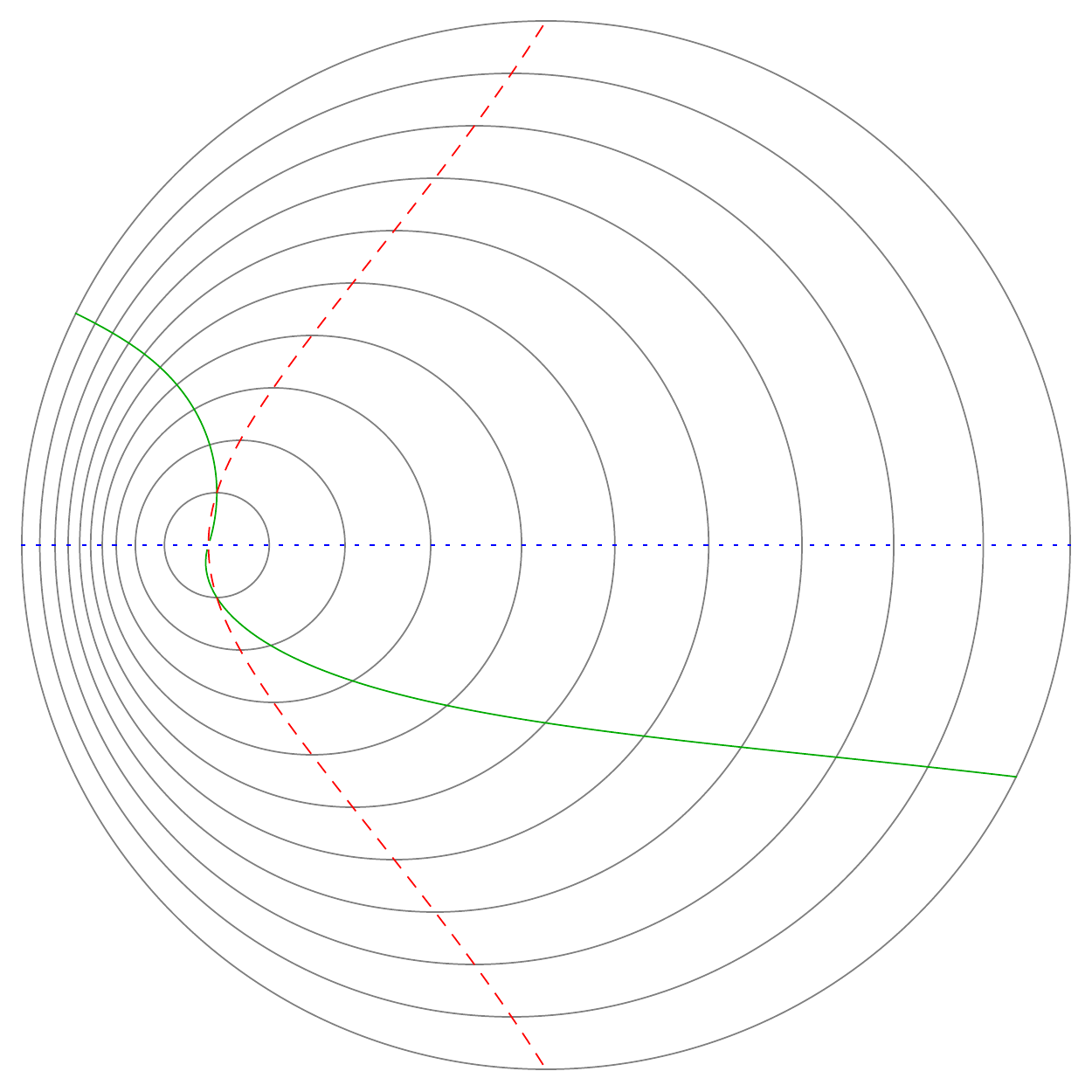}
\caption[Axial Drift]{An illustration of how $\theta_{\mathrm{ax}}$ changes with $r$ due to the shell rotation effect. The solid green line marks $\theta = 0$ on the top half and $\theta = \pi$ on the bottom half. The dashed red line shows where these points would be without the shell rotation effect, for comparison. The dotted blue line marks $\theta_{\mathrm{ax}}$.}
\label{fig:axialdrift}
\end{center}
\end{figure}

The angles $\theta_{\mathrm{ax}}$ cannot be constant because the shells rotate as $r$ increases, depending on the direction and magnitude of the dipolar asymmetry. Specifically,  
\begin{align}
\label{eq:axialdrift}
\theta_{\mathrm{ax}}' &= \left( \frac{E'}{E} \right)_{\mathrm{ax}} \sin \theta_{\mathrm{ax}}.
\end{align}
This means that as $r$ increases, any anisotropy along the axis rotates the shells' frames in a way that drives $\theta = 0$ towards the symmetry axis on the side where the shells are compressed, and $\theta = \pi$ towards the axis on the side where they are stretched apart, as shown in Fig.~\ref{fig:axialdrift}. With this illustration, we can see that even though the value of $\theta_{\mathrm{ax}}$ changes with $r$, it still traces a straight line across the model; the change in its value is due to the movement of the $\theta = 0$ reference point.

Despite appearing curved in spherical coordinates, though, the symmetry axis is indeed a straight line. These models can be indirectly obtained by starting with a model with only $S'$ nonvanishing and performing a Haantjes transformation, as we will describe in section~\ref{sec:Haantjes}.

\subsection{\label{sec:bilateralsymmetry}Bilateral symmetry}

A less restrictive form of symmetry is bilateral symmetry, in which case the directions of shell shifting and rotation all lie on the symmetry plane. The simplest way to achieve this is by setting $P'$ or $Q'$ to 0. In this case, the symmetry plane is $\phi = \pm \pi/2$ or $\phi = \left\{0,\pi \right\}$, respectively.

Slightly more generally, we can get bilateral symmetry by satisfying Eq.~(\ref{eq:axialsymmetrysol2a}), without Eq.~(\ref{eq:axialsymmetrysol2b}). This gives a symmetry plane passing through the poles and the point $(p_0,q_0)$, directed at angle $\phi = \arctan C_0$.

We may wish to place the symmetry plane off of the pole $\theta = 0$, though. This can be achieved by starting with a model satisfying Eq.~(\ref{eq:axialsymmetrysol2a}) and applying a Haantjes transformation, which is a rotation transformation we will detail in section~\ref{sec:Haantjes}. Remarkably, the end result matches the form of the equation for a great circle, Eq.~(\ref{eq:greatcircle}). That is, for any model, if there is a circle in the $(p,q)$ plane which, {\itshape for all $r$}, corresponds to a great circle on the sphere, then that model exhibits bilateral symmetry. Moreover, this circle marks the intersection of the symmetry plane with each shell.

\begin{figure*}[tbp]
\begin{center}
(a) \hspace{5.5cm} (b) \hspace{5.5cm} (c) \\
\raisebox{1cm}{\includegraphics[width=5.5cm]{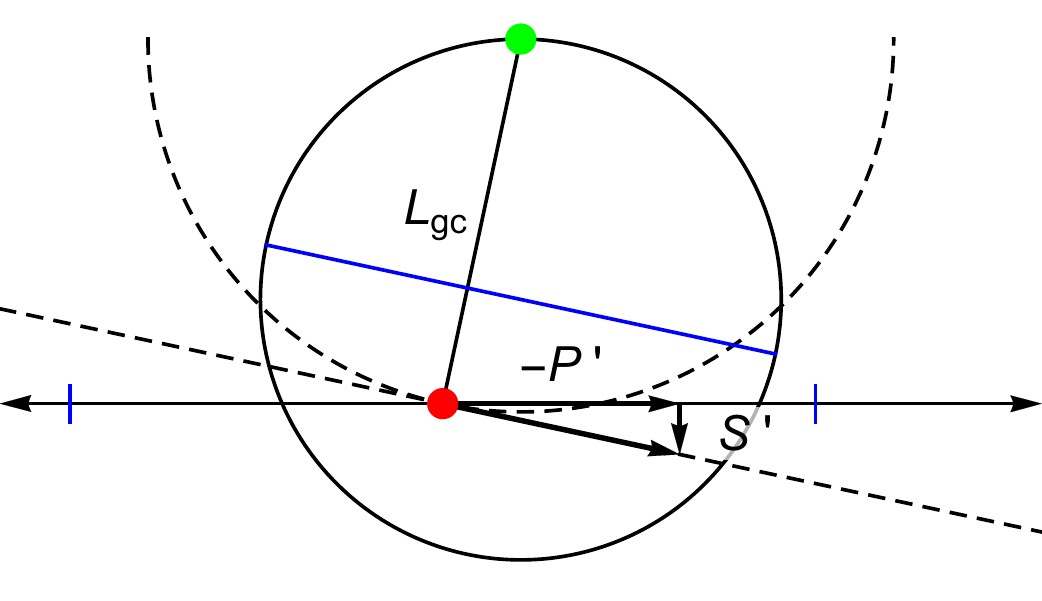}} \hspace{0.3cm}
\includegraphics[width=5.5cm]{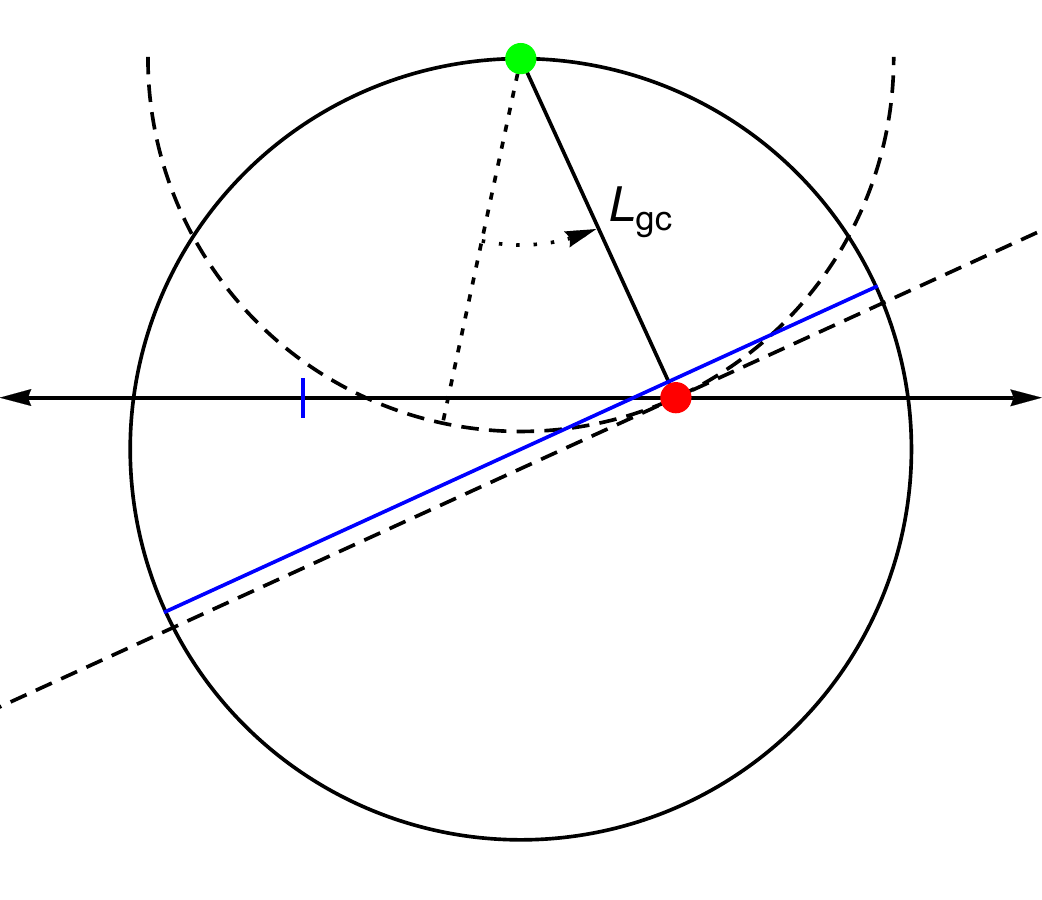} \hspace{0.3cm}  
\includegraphics[width=5.5cm]{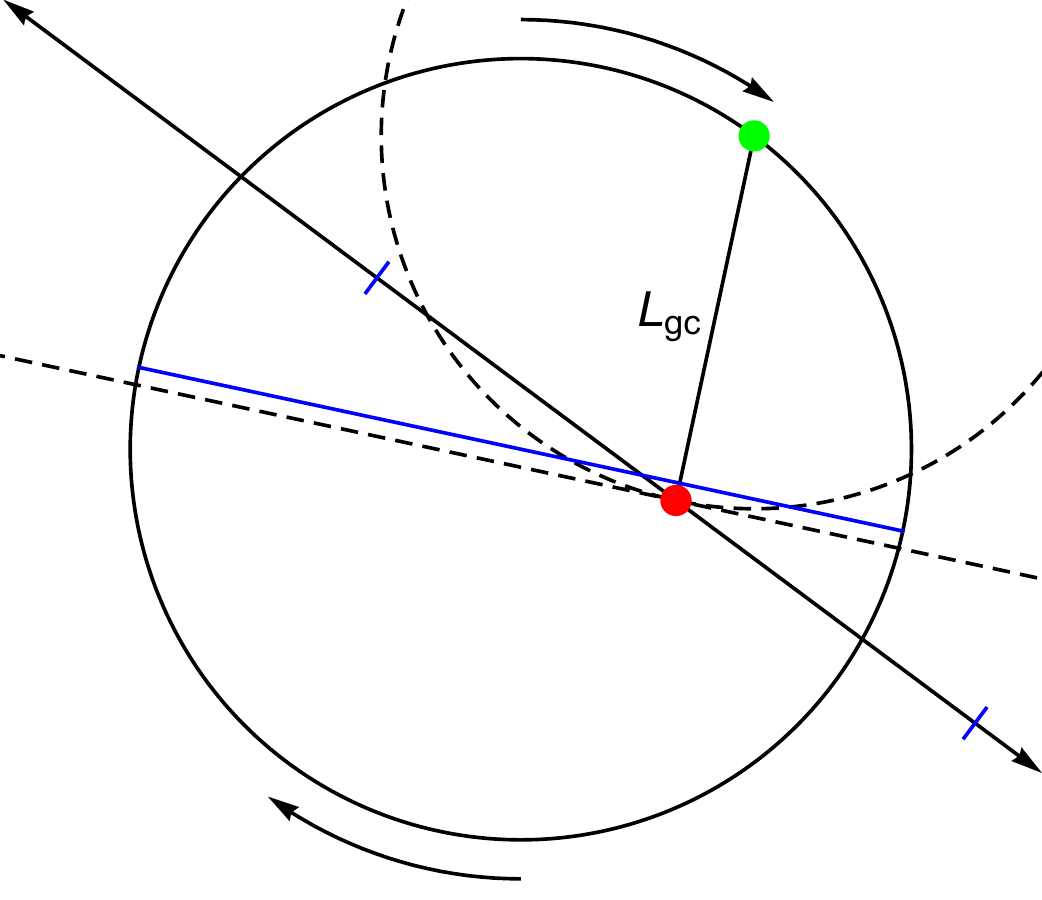}	
\caption[Bilateral Symmetry]{A side view of the situation described in the text, regarding the symmetry plane's compensating response to shell rotation. (a): The projection plane (solid black line with arrows) for a shell $r_1$ (solid circle) is positioned so that the point $p_c$ (red dot) is a distance $L_{\mathrm{gc}}$ from the projection point (green dot). The red dot can move as $r$ increases, but only in the directions tangent to the dashed circle. The total dipole direction is parallel to this motion, so it must lie on the great circle parallel to the dashed tangent line, shown edge-on as a solid blue line. The blue ticks on the projection plane mark where the edges of the great circle map to. (b): At a later shell $r_2$, the point $p_c$ has moved along the $L_{\mathrm{gc}}$ constraint sphere some distance. In the frame of shell $r_2$, it appears that the dipole lies on a different plane than before (i.e. the blue line is at a different angle). (c): The shell rotation effect rotates the whole picture, as indicated by the curved arrows. When we account for this, we see that the dipole plane does indeed coincide with that of shell $r_1$.}
\label{fig:bilateralsymmetry}
\end{center}
\end{figure*}

But how does such a simple constraint keep the Szekeres dipole pointing in the same plane, when the shells are being rotated in different directions? To develop an intuitive understanding of how this works, we first recall that the dipole functions determine the (negative) displacement of the projection plane and its origin relative to the projection point, as shown in Fig.~\ref{fig:projection}. Equation~(\ref{eq:greatcircle}), then, identifies $L_{\mathrm{gc}}$ as the distance between point $(p_c,q_c)$ and the projection point. If we hold $p_c$, $q_c$, and $L_{\mathrm{gc}}$ constant with $r$, we can imagine the two points to be connected by a rigid rod. We are allowed to move the projection plane around as $r$ increases, representing the dipole functions changing over $r$, but this motion is constrained to an imaginary sphere. This setup is illustrated in Fig.~\ref{fig:bilateralsymmetry}.

When we move the projection plane while moving outwards in $r$, we create a dipole anisotropy through the derivatives of the dipole functions. Because the motion of the plane is constrained, the direction of the dipole is constrained to the great circle parallel to the plane tangent to this imaginary sphere at the point $(p_c,q_c)$. This point is now at a different location on the sphere, so it would seem that the tangent plane has been rotated, but remember the rotation effect of the dipole functions---the shell rotates in a way that exactly counteracts the rotation of the tangent plane! With respect to a single shell, the tangent plane does not rotate, so all of the extrema lie in the same plane for all $r$, as long as Eq.~(\ref{eq:greatcircle}) holds. (The shell shifting effect is not a concern, because the shifts are all within the plane as well.)

This kind of symmetry does not confer most of the advantages of axial symmetry. It does not allow one to eliminate an entire coordinate from investigation; at best, it divides the number of directions that need to be considered by 2. It does allow for more complex arrangements of multiple structures than axially symmetric models, but for more than two structures it is too restrictive to be realistic. Perhaps the greatest advantage of bilateral symmetry is in graphical representation. Because paper and screens are two-dimensional, we can typically only clearly visualize a two-dimensional slice of the full model. If this slice is the symmetry plane, all of the structures will be neatly aligned within the page, making for a clear and faithful representation of the model. All of the figures in this paper showing examples of Szekeres models feature bilateral symmetry.

\section{\label{sec:methods}Methods and tools}

\subsection{\label{sec:tracking}Tracking shell shifting and rotation}

There are situations in which we need to be able to understand the geometry of a model, including its shell shifting and rotation, such as in generating spatially accurate images, as in Figs.~\ref{fig:Szekeffects} and \ref{fig:coarsemodel}. In axially symmetric models with $P$ and $Q$ constant, this is a straightforward matter of numerically calculating the shifts by integrating Eq.~(\ref{eq:zshift}). In general, though, care must be taken in calculating the orientations of each shell, as even the rotation axes change as the shells rotate. An example of a procedure for calculating the total displacements and rotations of each shell follows.

Begin at $r = 0$, with a $3 \times 3$ matrix $A$, set initially to the identity matrix. $A(r)$ represents the orientation of the LRF of shell $r$ relative to that of the innermost shells. That is, using the innermost shells' LRF as a basis, rows 1--3 of $A(r)$ give the components of the $x$, $y$, and $z$ unit vectors respectively of the LRF at shell $r$. Also create a three-component vector $\mathbf\Delta$ representing the total shifts, and initialize it to zero. Then, increment outwards in small steps of $\delta r$, up to a maximum $r_{\mathrm{max}}$, at each step doing the following:
\begin{enumerate}
	\item Calculate a new axis matrix 
	\begin{align}
	A(r + \delta r) = R_y\left(\frac{P'}{S} \; \delta r\right) \: R_x\left(-\frac{Q'}{S} \; \delta r\right) A(r),
	\end{align}
	where $R_x$ and $R_y$ are rotation matrices about the LRF's $x$ and $y$ axes,\footnote{The rotations approximately commute because their arguments are small.} defined as 
	\begin{align}
R_x(\psi) =  \begin{pmatrix}
1 & 0 & 0 \\
0 & \cos \psi & -\sin \psi \\
0 & \sin \psi & \cos \psi  \end{pmatrix} , \\
R_y(\psi) =  \begin{pmatrix}
\cos \psi & 0 & \sin \psi \\
0 & 1 & 0 \\
-\sin \psi & 0 & \cos \psi  \end{pmatrix} . 
\end{align}
	\item Calculate new shifts according to\footnote{As we will explain in the next subsection, it is sometimes appropriate to omit the $\sqrt{1-k(r)}$ factor here, depending on how one chooses to map the curved space onto a flat image.} 
	\begin{align}
	\label{eq:shiftarray}
	\mathbf{\Delta}(t,r + \delta r) = \mathbf{\Delta}(t,r) + \frac{R(t,r)}{\sqrt{1-k(r)}}A^T(r) \begin{pmatrix}
	P'/S \\ Q'/S \\ S'/S
	\end{pmatrix}(r).
	\end{align}
	\item Append the new shift value and axis matrix to an array, to keep track of both at each step in $r$.
\end{enumerate}
When finished, the array will hold all of the information about the displacement and orientation of shells up to $r_{\mathrm{max}}$. Note that the shifts calculated by this procedure are only valid at a single time slice. To plot at a different time, the shifts must be re-calculated. The axis matrix $\mathbf{A}(r)$, however, is invariant with time.

\subsection{\label{sec:plotting}Plotting}

Once we have the information about the shifts and rotations of each shell, we must still choose a mapping from the curved 3D space of the model to an image on flat 2D paper, if we wish to generate a plot like those in this paper. 

First of all, reducing the model to two dimensions is far simpler if the model has bilateral symmetry, as mentioned previously. The symmetry plane contains all of the extrema of the density function, and its intersection with any shell is a great circle, so the size of the circle drawn on the page corresponds neatly to the shell's size.

Next, we must eliminate the curvature. It is of course impossible to do so without introducing some kind of distortion. The simplest method is to pretend that $k(r) = 0$. This must be taken into account when building the shifts in Eq.~(\ref{eq:shiftarray}). We can then use the array of shifts and rotations from the previous subsection to map the Szekeres coordinates to Cartesian coordinates as
\begin{align}
\begin{pmatrix} X \\ Y \\ Z \end{pmatrix}(t, r, \theta, \phi) &=  
R(t,r) \, A^T(r) \begin{pmatrix} \sin \theta \cos \phi \\ \sin \theta \sin \phi \\ \cos \theta \end{pmatrix} + \mathbf{\Delta}(t,r).
\end{align}
(For 2D plots, we only use two of these coordinates.) With this mapping, each shell is drawn with radius $R(t,r)$. Distances along a shell's surface are conveyed accurately, but distances {\itshape between} shells are distorted, depending on the true curvature function.

Another choice of mapping preserves distances along lines of constant $(p,q)$, but distorts the sizes of shells and distances along their surfaces. The shells are drawn with a distorted radius 
\begin{align}
R_{map}(t,r) = \int_0^r{\frac{R'(t,\tilde{r})}{\sqrt{1-k(\tilde{r})}} \: \mathrm{d}\tilde{r}}
\end{align}
We build the shift array with the curvature factor included, and then map the model to Cartesian coordinates as
\begin{align}
\begin{pmatrix} X \\ Y \\ Z \end{pmatrix} &=
	R_{map}(t,r) A^T(r) \begin{pmatrix} \sin \theta \cos \phi \\ \sin \theta \sin \phi \\ \cos \theta \end{pmatrix} + \mathbf{\Delta}(t,r).
\end{align}

\subsection{\label{sec:geodesics}Calculating null geodesics}

The light beams we observe follow null geodesic paths, defined by
\begin{align}
\label{eq:nullgeodesics}
k^\alpha{}_{;\beta} k^\beta = 0,
\end{align}
where $k^\alpha = \mathrm{d}x^\alpha/\mathrm{d}\lambda$ is the tangent vector, and $\lambda$ is the affine parameter. The non-vanishing Christoffel symbols for the Szekeres metric are given in appendix~\ref{app:curvature}.

While the geodesic equations are fairly lengthy when fully written out, a simplification is possible through the definitions \cite{Nwankwo}
\begin{align}
\label{eq:F}
F &= \frac{R^2}{E^2}, \\
\label{eq:H}
H &= \frac{(R' - R\frac{E'}{E})^2}{1-k}.
\end{align}
With these compactified functions, the geodesic equations become 
\begin{subequations}
\label{eq:Szekeresgeodesics}
\begin{align}
\frac{\mathrm{d}k^t}{\mathrm{d}\lambda} + \frac{1}{2}H_{,t}(k^r)^2 + \frac{1}{2}F_{,t} \left[(k^p)^2 + (k^q)^2 \right] &= 0, \\
H \frac{\mathrm{d}k^r}{\mathrm{d}\lambda} + \frac{\mathrm{d}H}{\mathrm{d}\lambda} k^r - \frac{1}{2}H_{,r}(k^r)^2 - \frac{1}{2}F_{,r} \left[(k^p)^2 + (k^q)^2 \right] &= 0, \\
F \frac{\mathrm{d}k^p}{\mathrm{d}\lambda} - \frac{1}{2} H_{,p}(k^r)^2 + \frac{\mathrm{d}F}{\mathrm{d}\lambda} k^p - \frac{1}{2}F_{,p} \left[(k^p)^2 + (k^q)^2 \right] &= 0, \\
F \frac{\mathrm{d}k^q}{\mathrm{d}\lambda} - \frac{1}{2} H_{,q}(k^r)^2 + \frac{\mathrm{d}F}{\mathrm{d}\lambda} k^q - \frac{1}{2}F_{,q} \left[(k^p)^2 + (k^q)^2 \right] &= 0.
\end{align}
\end{subequations}
These equations can be numerically integrated to trace the path of a light beam through the model. A total of eight variables must be tracked (the position and tangent vector), using eight first-order differential equations---Eq.~(\ref{eq:Szekeresgeodesics}) and $k^\alpha = \mathrm{d}x^\alpha/\mathrm{d}\lambda$. Usually, we are interested in observed light beams, so we choose a location for the observer and a direction of observation to set the initial values for the eight variables, then propagate the equations backwards in time to the source.

The redshift along the beam is readily obtained by
\begin{align}
1+z &= \frac{(k_\alpha u^\alpha)_s}{(k_\alpha u^\alpha)_o},
\end{align}
where subscripts $s$ and $o$ denote the source and observer respectively, and $u^\alpha$ is the four-velocity of the source or observer. We can normalize the null geodesic tangent vector at the observer so that $k^t_o = -1$, and if we assume the source and observer are both comoving, we can set $u^\alpha_s = u^\alpha_o = (1,0,0,0)$, so we are left with simply 
\begin{align}
\label{eq:redshift}
1+z = -k^t_s.
\end{align}

In the case of axially symmetric models, there is a special case of geodesics which propagate along the symmetry axis, with $k^p = k^q = 0$ along their entire length. This simplifies the equations greatly. The null condition then gives a direct relation between $k^t$ and $k^r$, and thus between $t$ and $r$:
\begin{align}
\label{eq:axialnullcondition}
\frac{\mathrm{d}t}{\mathrm{d}r} &= \frac{k^t}{k^r} = \pm \frac{R' - R \, \frac{E'}{E}}{\sqrt{1-k}},
\end{align}
where the sign depends on which direction the geodesic is moving. Applying this to Eq.~(\ref{eq:Szekeresgeodesics}a), along with Eq.~(\ref{eq:redshift}), we can deduce an integral formula for redshift:
\begin{subequations}
\begin{align}
\label{eq:axialredshift}
\frac{\mathrm{d}(1+z)}{\mathrm{d}\lambda} &= \frac{\dot{R}' - \dot{R} \, \frac{E'}{E}}{R' - R \, \frac{E'}{E}} (1+z)^2,  \\
\frac{1}{1+z}\frac{\mathrm{d}(1+z)}{\mathrm{d}t_s} &= -\frac{\dot{R}' - \dot{R} \, \frac{E'}{E}}{R' - R \, \frac{E'}{E}},  \\
\ln (1+z) &= -\int_{t_o}^{t_s} \frac{\dot{R}' - \dot{R} \, \frac{E'}{E}}{R' - R \, \frac{E'}{E}} \: \mathrm{d}t,
\end{align}
\end{subequations}
where all quantities are evaluated on the geodesic, that is, with $r = r(t)$.

We can then express the geodesic equations directly in terms of $z$ \cite{BolejkoSC}:
\begin{subequations}
\label{eq:axialgeodesics}
\begin{align}
\frac{\mathrm{d}t}{\mathrm{d}z} &=  -\frac{1}{1+z} \frac{R' - R \, \frac{E'}{E}}{\dot{R}' - \dot{R} \, \frac{E'}{E}},	\\
\frac{\mathrm{d}r}{\mathrm{d}z} &= \pm \frac{1}{1+z} \frac{\sqrt{1-k}}{\dot{R}' - \dot{R} \, \frac{E'}{E}}.
\end{align}
\end{subequations}

\subsection{\label{sec:CoordinateTransformations}Coordinate transformations}

The labeling of coordinates in a Szekeres model has considerable flexibility. We have already mentioned the gauge freedom in the radial coordinate---defining a new coordinate $\widetilde{r} = f(r)$ (where $f$ is monotonic) gives a new description of the same physical model, obeying the same basic equations.

Likewise, the transverse coordinates can be transformed while maintaining the form of the metric---not as freely as the radial coordinate, but in several specific ways. 

\subsubsection{Translation}

The simplest transformation is a constant translation, 
\begin{align}
\label{eq:shiftcoords}
(\widetilde{p},\widetilde{q}) = (p + p_0, p + q_0).
\end{align}
To maintain the form of the metric, we also transform the dipole functions, as 
\begin{align}
\label{eq:shiftfunctions}
(\widetilde{P},\widetilde{Q},\widetilde{S}) = (P + p_0, Q + q_0, S).
\end{align}
This transformation only moves the origin point for the projective coordinate labeling; it does not affect the spherical coordinates. 

\subsubsection{Scaling}

It is also possible to perform a linear scaling transformation while maintaining the model's physical form. This is done by simply multiplying both projective coordinates and all three dipole functions by the same nonzero constant:
\begin{align}
\label{eq:scalecoords}
(\widetilde{p},\widetilde{q}) &= \mu (p, q), \\
\label{eq:scalefunctions}
(\widetilde{P},\widetilde{Q},\widetilde{S}) &= \mu (P,Q,S).
\end{align}
Again, this does not affect the spherical coordinates.

\subsubsection{Polar rotation}

A third simple transformation consists of rotating the $p$ and $q$ coordinates:
\begin{align}
(\widetilde{p},\widetilde{q}) &= (p \cos \psi + q \sin \psi, -p \sin \psi + q \cos \psi), \\
(\widetilde{P},\widetilde{Q},\widetilde{S}) &= (P \cos \psi + Q \sin \psi, -P \sin \psi + Q \cos \psi, S).
\end{align}

This {\itshape does} affect the spherical coordinates, by $\widetilde{\phi} = \phi \linebreak +\psi$---a simple rotation about $\theta = 0$ (with $\theta = \pi$ also fixed). While this axis is not in general the same for shells of different $r$, due to the shell rotation effect, the form of the metric and the physical arrangement of structures are preserved.

\subsubsection{Swapping $p$ and $q$}

We can easily see that the metric is invariant under the substitution
\begin{align}
\label{eq:swapcoords}
(\widetilde{p},\widetilde{q}) &= (q, p), \\
\label{eq:swapfunctions}
(\widetilde{P},\widetilde{Q}) &= (Q, P).
\end{align}
This amounts to a reflection across the $\phi = \pi/4$ plane. Combined with polar rotations, this can reflect the coordinates across any polar plane.

\subsubsection{Inversion}

As we saw in section~\ref{sec:coords}, the projective coordinates diverge where $\theta = 0$. This can cause problems for numerical calculations passing near this axis, even though there is nothing physically special happening there. When this occurs, a translation or scaling transformation cannot remove the infinity, but we can invert the coordinates so that the divergence occurs somewhere else, without changing the model's physical features or their positions relative to each other. This is done with the transformation
\begin{align}
(\widetilde{p},\widetilde{q}) &= \frac{(p,q)}{p^2 + q^2}, \\
\label{eq:inversionPQS}
(\widetilde{P},\widetilde{Q},\widetilde{S}) &= \frac{(P,Q,S)}{P^2 + Q^2 + S^2}.
\end{align}
Clearly, the point on any given shell where the projective coordinates previously diverged now has $\widetilde{p} = \widetilde{q} = 0$, and vice versa. Calculations can now proceed unimpeded through the region that was formerly problematic.

Because the angle $\theta = 0$ corresponds to the divergence point of the projective coordinates, and this point has moved (relative to physical structures), we can see that this transformation {\itshape does} affect the spherical coordinates. The point $\theta = 0$ corresponds to a new angle satisfying
\begin{align}
\cot \widetilde{\phi} = \frac{\widetilde{P}}{\widetilde{Q}}, \qquad
\cot \frac{\widetilde{\theta}}{2} = \frac{\sqrt{\widetilde{P}^2 + \widetilde{Q}^2}}{\widetilde{S}}.
\end{align}
The new point at which $\widetilde{\theta} = 0$ corresponds to an old angle in a similar fashion. More generally, the spherical and projective coordinates over the entire shell have been reflected across the great circle defined by 
\begin{align}
\label{eq:reflectionsurface}
p^2 + q^2 = P^2 + Q^2 + S^2,
\end{align}
which crosses the point halfway between $\theta = 0$ and ${p = q = 0}$. In spherical coordinates, this reflection surface is given by
\begin{align}
\label{eq:reflectionsurfacespherical}
P \sin \theta \cos \phi + Q \sin \theta \sin \phi + S \cos \theta = 0.
\end{align}
That is, in terms of the LRF, the reflection surface is a plane through the origin and perpendicular to the vector $(x,y,z)_{\mathrm{ref}} = (P,Q,S)/\sqrt{P^2+Q^2+S^2}$. It is then easy to see that we can adjust the orientation of this reflection surface by first performing a translation, as described above; such a translation simultaneously moves the point $p = q = 0$ and changes the LRF vector $(x,y,z)_{\mathrm{ref}}$. 

Note, however, that the definition of this reflection surface is a function of $r$. It is therefore not a flat plane across the entire model. Each shell's coordinates are reflected in a different direction, but the metric keeps the same form, and the overall picture still follows the same rules of shifting and rotation between shells laid out in section~\ref{sec:Efunction}, just with different dipole functions, Eq.~(\ref{eq:inversionPQS}). The examples of inversion shown in Fig.~\ref{fig:coordtransform} show how this works.

We can therefore avoid the $\theta = 0$ axis when calculating geodesics by performing this inversion operation whenever $p$ and $q$ get large enough to adversely affect precision. We must take care to also transform the geodesic tangent vector accordingly:
\begin{align}
(\widetilde{k^p},\widetilde{k^q}) = \frac{(k^p,k^q)}{p^2 + q^2} - 2(p,q) \frac{p k^p + q k^q}{(p^2 + q^2)^2}.
\end{align}

\subsubsection{\label{sec:Haantjes}Haantjes transformation}

A more general transformation allows us to rotate the coordinates about an arbitrary direction. This kind of transformation is called a Haantjes transformation \cite{Plebanski}.

In a quasispherical Szekeres model, a Haantjes transformation modifies the $p$ and $q$ coordinates as follows \cite{SitUbEM,PlebanskiBook}:
\begin{subequations}
\label{eq:Haantjes1}
\begin{align}
\widetilde{p} &= \frac{p + D_1 (p^2 + q^2)}{\tau}, \\
\widetilde{q} &= \frac{q + D_2 (p^2 + q^2)}{\tau}, \\
\tau &= 1 + 2 D_1 p + 2 D_2 q + (D_1^2 + D_2^2)(p^2 + q^2),
\end{align}
\end{subequations}
where $D_1$ and $D_2$ are arbitrary constants. In order to preserve the form of the metric, we must also transform the dipole functions:
\begin{subequations}
\label{eq:Haantjes2}
\begin{align}
\widetilde{P} &= \frac{P + D_1 (P^2 + Q^2 + S^2)}{T}, \\
\widetilde{Q} &= \frac{Q + D_2 (P^2 + Q^2 + S^2)}{T}, \\
\widetilde{S} &= \frac{S}{T}, \\
T &= 1 + 2 D_1 P + 2 D_2 Q + (D_1^2 + D_2^2)(P^2 + Q^2 + S^2).
\end{align}
\end{subequations}

This transformation amounts to a rotation about the points satisfying
\begin{subequations}
\begin{align}
p^2 + q^2 &= P^2 + Q^2 + S^2,  \\
D_1 p + D_2 q &= D_1 P + D_2 Q,
\end{align}
\end{subequations}
by an angle of 
\begin{align}
\psi = 2 \arccos \left[ \frac{(P,Q,S) \cdot (P + D_1, Q + D_2, S)}{\left\Vert{(P,Q,S)}\right\Vert \left\Vert{(P + D_1, Q + D_2, S)}\right\Vert} \right],
\end{align}
where $\left\Vert{(P,Q,S)}\right\Vert = \sqrt{P^2+Q^2+S^2}$. The rotation axis is restricted to the great circle halfway between $\theta = 0$ and $p = q = 0$, but this great circle can be moved by first performing a coordinate translation. By exercising this freedom in combination with the free variables $D_1$ and $D_2$, the rotation can be of any angle about any axis.

\begin{figure*}[tbp]
\begin{center}
(a) \hspace{4cm} (b) \hspace{4cm} (c) \hspace{4cm} (d) \\
\includegraphics[width=3.8cm]{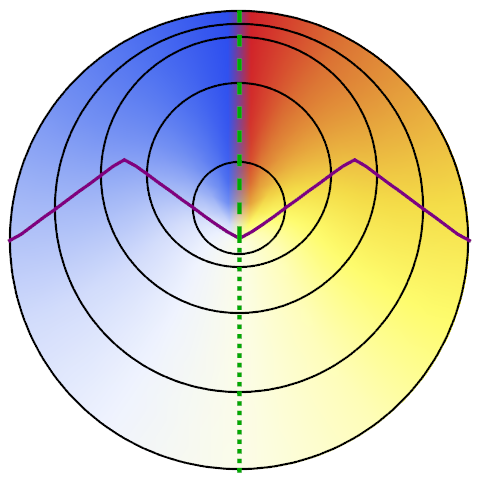} \hspace{0.6cm}
\includegraphics[width=3.8cm]{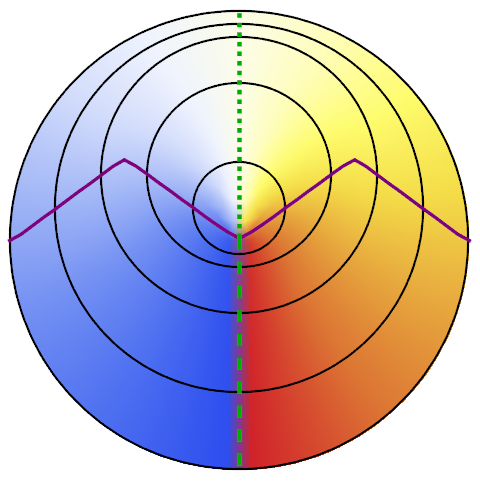} \hspace{0.6cm}  
\includegraphics[width=3.8cm]{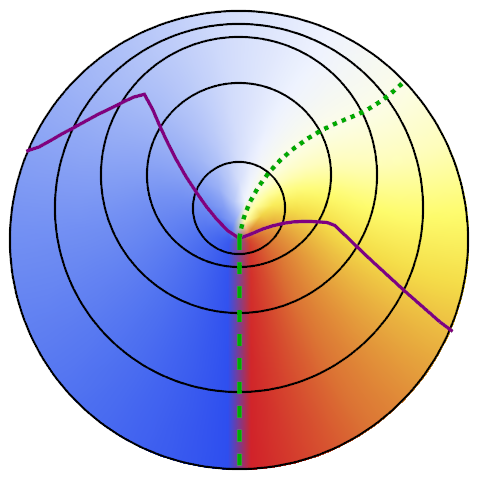}	\hspace{0.6cm}  
\includegraphics[width=3.8cm]{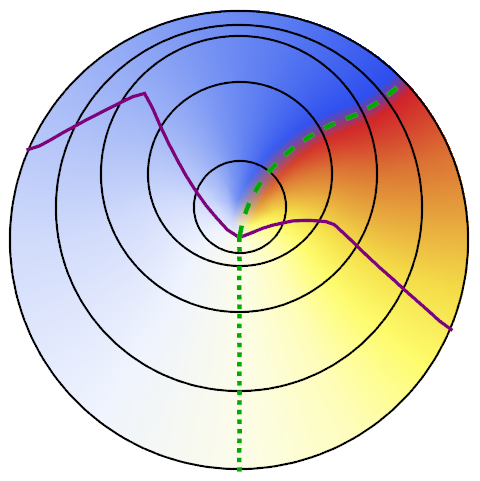}
\caption[Coordinate Transformations]{Visualization of a series of coordinate transformations. Colors indicate the $p$ coordinate, from blue at negative infinity to red at positive infinity ($q$ is 0 in the displayed slice), and the dotted and dashed green lines show where it vanishes and diverges, respectively (the LRF's $-z$ and $+z$ directions). The purple line is the reflection surface defined in Eq.~(\ref{eq:reflectionsurface}). (a): the initial state, an axially symmetric model with $P(r) = Q(r) = 0$. It then undergoes an inversion transformation, with the result shown in (b). (c) shows the effect of a translation of the $p$ coordinate, and (d) is after a second inversion. The combined effect is a Haantjes transformation.}
\label{fig:coordtransform}
\end{center}
\end{figure*}

The Haantjes transformation can be seen as a combination of three simpler transformations: an inversion, followed by a translation of $(D_1, D_2)$, followed by another inversion. The first inversion reflects the coordinates across the surface $p^2 + q^2 = P^2 + Q^2 + S^2$, and the translation reorients this surface before the second inversion reflects the coordinates again. This sequence is illustrated in Fig.~\ref{fig:coordtransform}.

Haantjes transformations can be useful in model construction, as they provide a means of moving an axially symmetric anisotropy to any desired direction. One can create a model with $P = Q = 0$, meaning anisotropies only arise in the $z$ direction as a result of $S$, and then one can apply an appropriate Haantjes transformation to reorient the anisotropy to any direction, with a mixture of all three dipole functions automatically satisfying the conditions for axial symmetry, Eq.~(\ref{eq:axialsymmetry}). The $(D_1, D_2)$ values needed to transform the axis to the direction $(\theta,\phi)$ (measured at the lowest $r$ value of the anisotropy, $r_l$) are given by 
\begin{subequations}
\label{eq:Haantjes3}
\begin{align}
D_{1} &= \frac{1 - \cos \theta}{\sin \theta}\cos \phi \: e^{-S(r_l)}, \\ 
D_{2} &= \frac{1 - \cos \theta}{\sin \theta}\sin \phi \: e^{-S(r_l)}.
\end{align}
\end{subequations}
Due to the shell rotation effect, the $(\theta,\phi)$ values of the maximum anisotropy will not be constant with $r$ after the transformation. The $D$ parameters are related to the $C$ coefficients of the axial symmetry equations (\ref{eq:axialsymmetrysol2}) by 
\begin{subequations}
\label{eq:Haantjessymmetryrelation}
\begin{align}
C_0 &= \frac{D_2}{D_1},	\\
C_1 &= \frac{1}{D_1}.
\end{align}
\end{subequations}
The new symmetry axis has coordinates
\begin{subequations}
\label{eq:Haantjessymmetryaxis}
\begin{align}
p_0 &= \frac{D_1}{D_1^2 + D_2^2},	\\
q_0 &= \frac{D_2}{D_1^2 + D_2^2},
\end{align}
\end{subequations}
on one side, and $(0,0)$ on the other.

\subsection{\label{sec:construction}Example construction method: randomized series of structures}

One of the key strengths of the Szekeres metric is its ability to simulate multiple structures. When designing such a model, though, we should be careful of how the shell rotation effect influences the structures' shapes. As we have seen in Fig.~\ref{fig:Szekeffects}, overdensities that may appear symmetrical (i.e., with the ratios of $P'/S$, $Q'/S$, and $S'/S$ held constant over some range of $r$) are in fact smeared by the shell rotation, if they are not centered at $\theta = 0$ or $\pi$. This can systematically distort our structures based on their orientation, which is undesirable if we wish them to have similar shapes. 

We have devised a process for generating a randomized series of structures that are all the same kind of shape---individually axially symmetric---regardless of their orientation. This method involves performing Haantjes transformations in a piecewise manner over separate intervals of $r$. Because we do not apply the transformations globally, they are no longer simply coordinate transformations, but physical rearrangements of structures. The process goes as follows: 
\begin{enumerate}
	\item First, we divide the model into separate intervals of $r$, with boundaries $r_i$, with $i = 0..n$, $r_0 = 0$, and $r_{i+1} > r_i$. We introduce axially symmetric Szekeres anisotropies by making $S$ a piecewise function, with $S'/S$ vanishing at all $r_i$ (to ensure continuity when we are done). 
	\item Then, we generate a list of random angles $(\theta_i, \phi_i)$, which will correspond to the directions of the maximum density contrast at the lower bound of each section. To move the anisotropies to these randomly chosen angles, we apply the Haantjes transformation separately to each interval, using Eq.~(\ref{eq:Haantjes2}). The $D$ coefficients for each interval are given by Eq.~(\ref{eq:Haantjes3}), using $(\theta_i, \phi_i)$ for the interval $r_i < r < r_{i + 1}$.
	\item The dipole functions will now be discontinuous at each $r_i$, creating shell crossing singularities. To remedy this, we first address the discontinuities in $S$ using a scaling transformation, Eq.~(\ref{eq:scalefunctions}). One interval at a time, starting with $i = 1$, we scale the dipole functions by a factor $S_{i-1}(r_i)/S_{i}(r_i)$ (where $S_{i}$ is the $S$ function in the interval $r_i < r < r_{i + 1}$).
	\item The $S$ function is now continuous, but the $P$ and $Q$ functions are not. This can be fixed similarly with a series of shifts, using Eq.~(\ref{eq:shiftfunctions}). Again starting at $i = 1$, we shift the $p$ coordinate by $P_{i-1}(r_i) - P_{i}(r_i)$, and likewise for $q$, and proceed through the remaining intervals one at a time.
\end{enumerate}

\begin{figure}[tbp]
\begin{center}
\includegraphics[width=8cm]{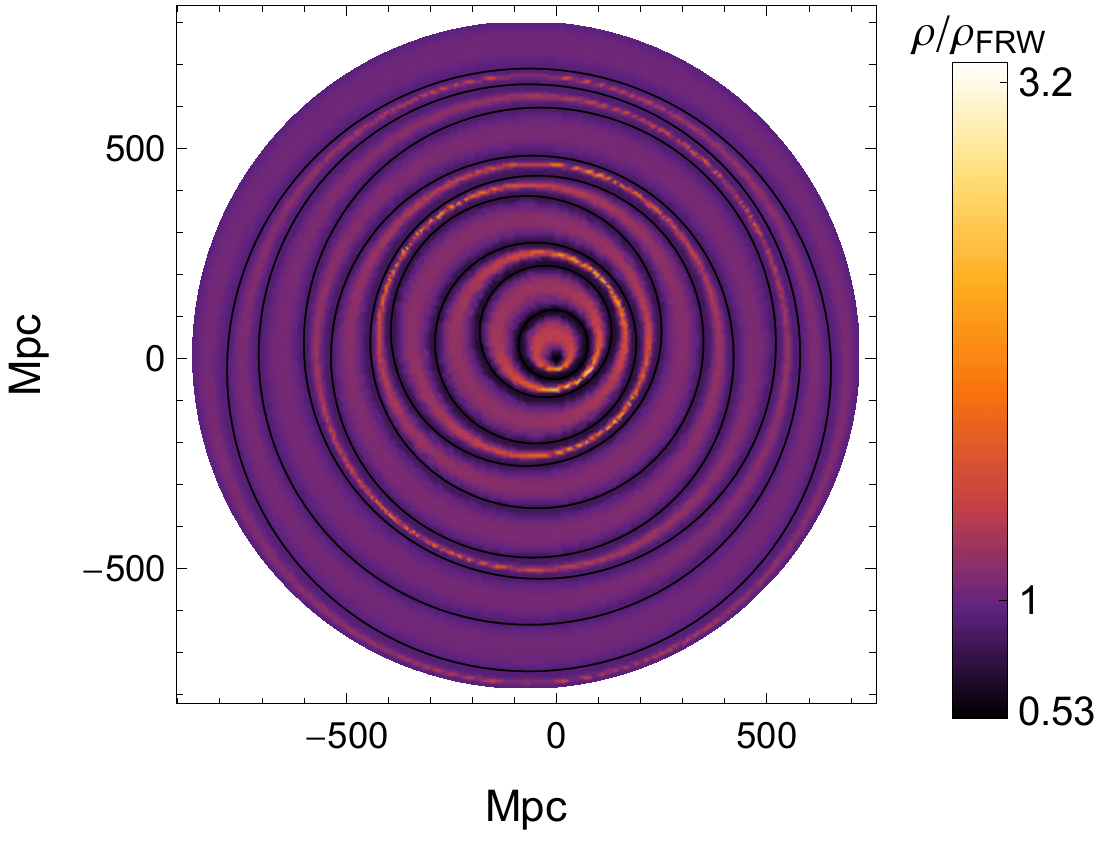} 
\caption[Coarse model]{An example of a model constructed through randomized piecewise Haantjes transformations. The structure angles were chosen over the unit circle rather than the unit sphere, so that the model maintains a symmetry plane on which all of the maximum density contrasts fall, making for a better picture, but the construction process is fully capable of creating three-dimensional arrangements.}
\label{fig:coarsemodel}
\end{center}
\end{figure}

We are left with a single model containing an arrangement of structures, each individually axially symmetric, but together completely asymmetric in a randomized fashion. An example is illustrated in Fig.~\ref{fig:coarsemodel}. Each structure is oriented in the direction of $(\theta_i, \phi_i)$ {\itshape at the lower boundary of the r interval}; due to shell rotation within the interval, the angle will generally be different at the upper boundary, following Eq.~(\ref{eq:axialdrift}). Nevertheless, if the angles were chosen uniformly over the sphere, there will be no correlation between the orientations of sequential structures.

This method of model construction results in a coarse-grained array of large, pancake-like structures. This can be done with any base LT model, for instance creating a series of overdense walls inside a giant void, as in \cite{BolejkoCoarse}. But it fits particularly well with a model with a radially oscillating density, following the same series of $r$ intervals. These kinds of models, called ``Onion'' models by some \cite{BiswasOnion,KoksbangRayTrace}, are compatible with Sussman's prescription of periodic local homogeneity (PLH) \cite{Sussman}.

\section{\label{sec:embedding}Embedding in 4-dimensional space}

As we mentioned previously, the Szekeres model's curved geometry makes it impossible to depict directly on flat paper without distortion. It is possible, though, to view a const-$t$ slice as a curved hypersurface embedded within a 4-dimensional space. With such a description, we can examine unambiguously how the curvature function and dipole functions interact to produce the Szekeres metric. This also gives confirmation that the shell shifting and rotation effects we have described do indeed fully account for the form of the metric.

If the curvature function is everywhere non-negative, an embedding is possible within a flat Euclidean background manifold. If the curvature function is negative anywhere, we can use a background manifold with constant negative curvature. We consider these two cases separately.

\subsection{Non-negative curvature}

Let the background space have coordinates $(X,Y,Z,W)$, with a line element $\mathrm{d}s^2 = \mathrm{d}X^2 + \mathrm{d}Y^2 + \mathrm{d}Z^2 + \mathrm{d}W^2$. The origin of the Szekeres model hypersurface coincides with the origin of the $(X,Y,Z,W)$ coordinates, and it extends initially outwards in the $X$-$Y$-$Z$ hyperplane, which is aligned with the LRF's $x$, $y$, and $z$ axes. We can then identify $(X,Y,Z,W) \approx R(t_0,r)(\sin \theta \cos \phi, \sin \theta \sin \phi, \cos \theta, 0)$ for sufficiently small $r$. (Because we are only looking at one spatial slice at a time, we will use $t = t_0$ everywhere.)

In an LT model, a positive curvature function gives the surface a bowl shape, rising in the $W$ direction as $r$ increases. Each shell $r$ is a 2-sphere of radius $R$, contained in a hyperplane parallel to the the $X$--$Y$--$Z$ hyperplane. The slope of the bowl is given by 
\begin{align}
\label{eq:EmbedAlpha}
\alpha_w(r) &= \pm \sqrt{\frac{k(r)}{1-k(r)}}.
\end{align}
This is what gives radial distances the extra curvature factor. Going from $r$ to $r + \mathrm{d}r$, we must move outwards by $R' \; \mathrm{d}r$, but also ``upwards'' in the orthogonal $W$ direction by $\alpha_w \, R' \; \mathrm{d}r$, for a total distance of $\sqrt{1 + \alpha_w^2} \; R' \; \mathrm{d}r$, or $R' \; \mathrm{d}r/\sqrt{1-k}$.

\begin{figure}[tbp]
\begin{center}
(a)\\ \includegraphics[width=8cm]{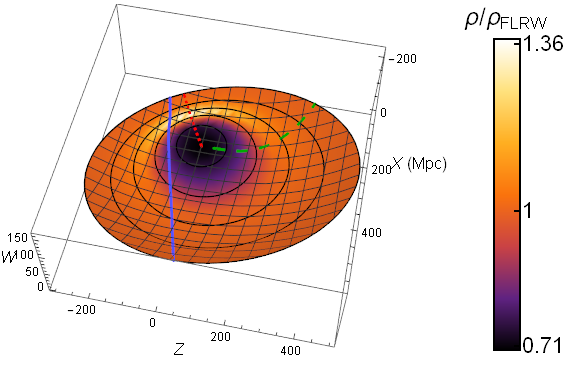}	\\
(b)\\ \includegraphics[width=8cm]{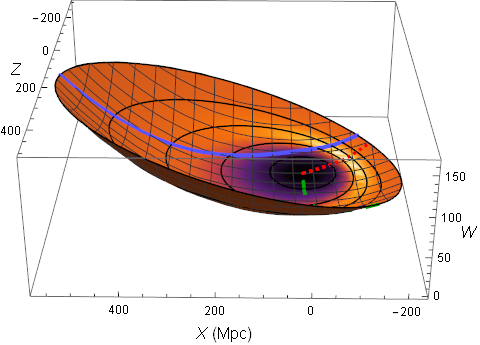}	\\
(c)\\ \includegraphics[width=8cm]{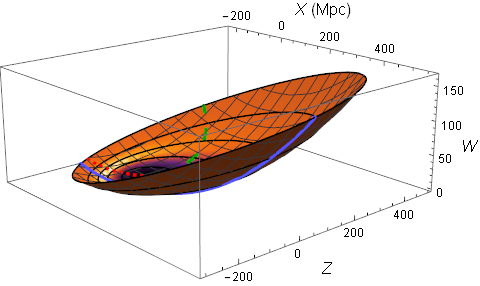}	
\caption[PositiveEmbed]{Three views of a slice of a positively-curved Szekeres model as an embedding in a 4-dimensional Euclidean manifold. The blue line shows the path of an arbitrary geodesic. Black circles mark shells of constant $r$ in steps of $80$ Mpc. The green dashed line marks $\theta = 0$, and the red dotted line marks $\theta_{\mathrm{max}}$, where the density has its angular maximum. Model definitions are given in appendix~\ref{app:example}.}
\label{fig:embedding}
\end{center}
\end{figure}

Positively curved Szekeres models have the same sort of bowl geometry, but with a modification due to the dipole functions. We have seen how the dipole functions shift the shells relative to each other, so over a span $\delta r$, the bowl extends farther outwards on one side than the other. But because the curvature function in the metric modifies the entire term $R' - R \: E'/E$, not just $R'$, the slope of the bowl must be the same everywhere around the shell.\footnote{This slope is measured in terms of the LRF, not the background coordinates.} This means that, from the viewpoint of this embedding, the dipole functions have an extra geometric effect not previously discussed: they not only shift and rotate shells, but also {\itshape tilt} them within the background manifold. Specifically, in terms of the LRF at shell $r$ (which now has a fourth direction, ``$w$'', orthogonal to the shell's hyperplane), the shell at $r + \mathrm{d}r$ makes an angle $\frac{P'}{S} \alpha_w \: \mathrm{d}r$ in the $x$--$w$ plane, $\frac{Q'}{S}\alpha_w \: \mathrm{d}r$ in the $y$--$w$ plane, and $\frac{S'}{S}\alpha_w \: \mathrm{d}r$ in the $z$--$w$ plane. Because of this tilting, the ``upwards'' shifts due to the bowl's slope do not go in the background's $W$ direction, but rather the LRF's $w$ direction.\footnote{This is why we have called the slope $\alpha_w$ instead of $\alpha_W$.} The resulting shape is illustrated in Fig.~\ref{fig:embedding}.

The approach for determining the orientation of each shell is similar to the one presented in section~\ref{sec:tracking}, with the addition of another dimension. We define a $4 \times 4$ matrix $A(r)$ that relates a shell's LRF axes to the axes of the background manifold (with rows 1--4 giving the $x$, $y$, $z$, and $w$ directions respectively). It is initialized as $A_{ij}(0) = \delta_{ij}$ for simplicity.\footnote{Any initialization is acceptable, as long as it is orthonormal.} As $r$ increases, it evolves according to	
\begin{align}
\label{eq:embedaxesevolution}
A'(r) &= \begin{pmatrix} 
0	&	0	&	\frac{P'}{S}	&	\frac{P'}{S}\alpha_w	\\
0	&	0	&	\frac{Q'}{S}	&	\frac{Q'}{S}\alpha_w	\\
-\frac{P'}{S}	&	-\frac{Q'}{S}	&	0	&	\frac{S'}{S}\alpha_w	\\
-\frac{P'}{S}\alpha_w	&	-\frac{Q'}{S}\alpha_w	&	-\frac{S'}{S}\alpha_w	&	0	\\
\end{pmatrix}A(r).
\end{align}

The Szekeres model's hypersurface, then, is defined by
\begin{align}
\label{eq:embedsurface}
\begin{pmatrix}
	X \\ Y \\ Z \\ W
	\end{pmatrix}(t_0,r) &= R(t_0,r) \; A^T(r) \begin{pmatrix}
	\sin \theta \cos \phi \\ \sin \theta \sin \phi \\ \cos \theta \\ 0
	\end{pmatrix} + \mathbf{\Delta}(t_0,r),
\end{align}
where $\mathbf{\Delta} = \mathbf{\Delta}(t_0,r)$ is a 4-component vector denoting the total displacement of the center of shell $r$ relative to the origin. It satisfies 
\begin{align}
\label{eq:embedshifts}
\mathbf{\Delta}'(t_0,r) &= A^T(r) \begin{pmatrix}
	R \; P'/S \\ R \; Q'/S \\ R \; S'/S \\ R' \; \alpha_w
	\end{pmatrix}(t_0,r).
\end{align}
We initialize it as $\Delta_i(t_0,0) = 0$, for simplicity. Note that the shifts from the dipole functions do not have the curvature factor included in Eqs.~(\ref{eq:xshift})--(\ref{eq:zshift}), as that part of the displacement is accounted for in the fourth component.

\subsection{Arbitrary curvature}

If $k(r) < 0$ for some $r$, the quantity $\alpha_w$ defined in Eq.~(\ref{eq:EmbedAlpha}) is not real, so an embedding in a 4-dimensional Euclidean manifold as above is not possible. We can, however, adapt the above embedding scheme for a 4-dimensional background manifold of constant negative curvature.

The line element of the background manifold can be written in terms of a radial coordinate $\rho$ and three angular coordinates $(\alpha, \beta, \gamma)$ as
\begin{align}
\label{eq:NegCurveMetric}
\mathrm{d}s^2 = &\frac{1}{1 - k_b \rho^2} \mathrm{d}\rho^2 \nonumber \\
&+ \rho^2 \left(\mathrm{d}\alpha^2 + \sin^2 \alpha \:\mathrm{d}\beta^2 + \sin^2 \alpha \: \sin^2 \beta \: \mathrm{d}\gamma^2\right), 
\end{align}
where $k_b$ is the curvature of the background. We will use $\chi(t,r,\theta,\phi)$ to denote the set of background coordinates $(\rho,\alpha,\beta,\gamma)$ corresponding to a point on the Szekeres hypersurface.

We consider the region local to a particular shell $r_0$. We align the background coordinates to this shell, so that
\begin{align}
\label{eq:NegCurveEmbedAlignment}
\chi(t_0,r_0, \theta, \phi) = \left(R, \frac\pi2, \theta, \phi \right).
\end{align}
The only parts of the embedding in the non-negative curvature case that depended on the curvature were the shifts and tilts in the local $w$ direction, through $\alpha_w$. We will assume for now that everything works as before, and we only need to find the new $\alpha_w$. For this, it is sufficient to consider a model in which $P' = Q' = 0$, so we only need to deal with $S$.

The shell at $r_0 + \mathrm{d}r$ has coordinates in the background manifold
\begin{widetext}
\begin{align}
\label{eq:NegCurveEmbed0}
\chi(t_0,r_0 + \mathrm{d}r, \theta, \phi)&= \chi(t_0,r_0,\theta,\phi) + \left[R' + R \frac{S'}{S} \cos \theta, -\alpha_w \left( \frac{R'}{R} + \frac{S'}{S} \cos \theta \right), -\frac{S'}{S} \sin \theta, 0 \right]\mathrm{d}r.
\end{align}
By taking $\chi(t_0,r_0 + \mathrm{d}r, \theta_0 + \mathrm{d}\theta, \phi_0 + \mathrm{d}\phi) - \chi(t_0,r_0, \theta_0, \phi_0)$ and plugging the results into the background metric, Eq.~(\ref{eq:NegCurveMetric}), we can obtain the metric on the Szekeres hypersurface:
\begin{align}
\label{eq:NegCurveHypersurface}
\mathrm{d}s^2 &= \left[ \frac{\left(R' + R \frac{S'}{S} \cos \theta \right)^2}{1 - k_b R^2} + R^2 \frac{S'^2}{S^2} \sin^2 \theta + \alpha_w^2 \left(R' + R \frac{S'}{S} \cos \theta \right)^2 \right] \mathrm{d}r^2 - 2R^2 \frac{S'}{S} \sin \theta \: \mathrm{d}r \, \mathrm{d}\theta + R^2 \mathrm{d}\Omega^2.
\end{align}
\end{widetext}
From here, we can compare to the Szekeres metric in spherical coordinates, Eq.~(\ref{eq:sphericalmetric}). Requiring $g_{rr}$ to match gives us an equation we can solve for $\alpha_w$:
\begin{align}
\label{eq:NegCurveAlphaSolution}
\frac{1}{1 - k_b R^2} + \alpha_w^2 &= \frac{1}{1 - k} \nonumber \\
\alpha_w = \pm \sqrt{\frac{1}{1-k} - \frac{1}{1 - k_b R^2}} &= \pm \sqrt{\frac{k - k_b R^2}{(1-k)(1 - k_b R^2)}}
\end{align}
This solution is only real if $k_b R^2 < k$. This means that we require the background curvature to be more negative than the strongest negative curvature of the Szekeres model.

In addition to the background manifold being non-Euclidean, note also that $\alpha_w$ depends on $R$, which is a function of $t$. Unlike the flat-background case, here the slope of the embedding surface changes over time.

With this solution for $\alpha_w$, the local metric around the shell $r_0$ matches the Szekeres metric, as desired. We can envision extending this embedding for the entire Szekeres model by moving both inwards and outwards in incremental steps $\mathrm{d}r$, at each step reorienting the background coordinates to match Eq.~(\ref{eq:NegCurveEmbedAlignment}). We then have a global embedding, though it is not expressed as concisely as in Eq.~(\ref{eq:embedsurface}). We can also confirm that this method works when $P'$ and $Q'$ are non-zero, though the equations get considerably more lengthy.

\section{\label{sec:discussion}Discussion}

Quasispherical Szekeres models are an anisotropic generalization to the spherically symmetric Lema\^itre-Tolman (LT) models. They have great potential as fully general-relativistic representations of cosmic structures. While they are well-understood mathematically, their structures are somewhat opaque to intuition. It is easy to misconstrue the physical picture behind the equations.

The dipole functions alter the matter distribution and the geometry of the models in multiple ways, which we have thoroughly explained. The shells of constant $t$ and $r$ are not only non-concentric, but also non-aligned. This relative rotation of the shells has often been overlooked, but as we have shown, it has deep connections to several other aspects of the model, such as the relationship between the two most common coordinate systems (projective and spherical), the conditions for axial symmetry, and geodesic paths. With this shell rotation effect properly accounted for, we can generate density plots that show the true shape of the model more accurately than before. We see that geodesics appear nearly straight, as they should, while structures defined by simple functions are sometimes skewed in ways not immediately obvious.

The conditions by which Szekeres models are axially symmetric are well-known, but with a better understanding of the overall geometry dictated by the model functions, we can understand the equations more intuitively. Moreover, we have shown that a less restrictive condition results in bilateral symmetry, which we have also illustrated in terms of the shell rotation effect. We have also reviewed the coordinate transformations that preserve the form of the metric, and used these in combination with the symmetry conditions to give an example of a model construction method that produces a randomized series of simple structures.

Finally, we have used our understanding of the geometry to build the spatial Szekeres metric as a 3-dimensional surface embedded in a 4-dimensional space. Because the metric on this surface fully matches the Szekeres metric, this confirms that the effects we have described tell the whole story. That is, the shell rotation effect is real, and there are no other hidden geometric effects waiting to be discovered. 

We have sought to provide the reader with a firm understanding of the Szekeres models' basic properties, as well as some of the practical tools needed to work with them. This is not a comprehensive analysis of the properties of the Szekeres models. Other works have already explored certain aspects of them in greater detail. There is a great deal of potential work yet to be done, though, and it is important to hold a clear picture of what we are working with as we forge ahead.

\begin{acknowledgements}
We would like to thank Dr. Charles Hellaby for very helpful comments and discussion. Funding for this research was provided by the Vaughan Family Endowment. Graphics were generated and some computations were performed with the Wolfram Mathematica 10 software \cite{Mathematica}.
\end{acknowledgements}

\appendix

\section{\label{app:solutions}Solutions to the evolution equations}

In the case of $\Lambda = 0$, the solution to Eq.~(\ref{eq:evolution}) is explicit in a parametric form, using a parameter $\eta(t,r)$:
\begin{align}
\label{eq:L0sols}
R(t,r) &= \begin{cases}
		\frac{M}{k}(-1 + \cosh \eta) & \text{if } k(r) < 0, \\
		M \: \eta^2/2 & \text{if } k(r) = 0, \\
		\frac{M}{k}(1 - \cos \eta) & \text{if } k(r) > 0,
	\end{cases} \\
t - t_B(r) &= \begin{cases}
		\frac{M}{(-k)^{3/2}}(-\eta + \sinh \eta) & \text{if } k(r) < 0, \\
		M \: \eta^3/6 & \text{if } k(r) = 0, \\
		\frac{M}{k^{3/2}}(\eta - \sin \eta) & \text{if } k(r) > 0.
	\end{cases} 
\end{align}
These cases are hyperbolic, parabolic, and elliptic, respectively. In practice, the precision of a numerically calculated solution suffers near points where $k(r)$ crosses $0$, so it is sometimes necessary to use a series expansion to handle ``near-parabolic'' regions (see also \cite{HellabyLTstructure}, app. B):
\begin{align}
\label{eq:nearparabolic}
R(t,r) &= M \left( \frac{\eta^2}{2} - \frac{\eta^4}{24}k + \frac{\eta^6}{720}k^2 - \frac{\eta^8}{40320}k^3 + \cdots \right), \\
t - t_B(r) &= M \left( \frac{\eta^3}{6} - \frac{\eta^5}{120}k + \frac{\eta^7}{5040}k^2 - \frac{\eta^9}{362880}k^3 + \cdots \right) .
\end{align}
The precise conditions for this solution to apply depend on the model details, gauge choices, and  code structure. One example condition is $\lvert k(r) \rvert < \zeta_0 \lvert k'(r) \rvert r$, where $\zeta_0$ is a small constant.

If $\Lambda \neq 0$, the solution requires an elliptic integral arising from Eq.~(\ref{eq:bangtime}).

\section{\label{app:curvature}Christoffel symbols and curvature tensors}

The non-zero Christoffel symbols for the quasispherical Szekeres metric are
\begin{widetext}
\begin{align}
\label{eq:Christoffel}
\Gamma^t{}_{rr} &= \frac{(R_{,r} - R \, \frac{E_{,r}}{E})(R_{,tr} - R_{,t} \, \frac{E_{,r}}{E})}{1-k}	& 
\Gamma^p{}_{rr} &= \frac{(R_{,r}/R - E_{,r}/E)(E_{,rp} E - E_{,r} E_{,p})}{1 - k}	\nonumber	\\
\Gamma^t{}_{pp} &= \Gamma^t{}_{qq} = \frac{R_{,t} \: R}{E^2} & 
\Gamma^q{}_{rr} &= \frac{(R_{,r}/R - E_{,r}/E)(E_{,rq} E - E_{,r} E_{,q})}{1 - k}	\nonumber	\\
\Gamma^r{}_{rt} &= \Gamma^r{}_{tr} = \frac{R_{,tr} - R_{,t} \, \frac{E_{,r}}{E}}{R_{,r} - R \, \frac{E_{,r}}{E}}	&
\Gamma^p{}_{pp} &= -\Gamma^p{}_{qq} = \Gamma^q{}_{pq} = \Gamma^q{}_{qp} = -\frac{E_{,p}}{E} \nonumber	\\
\Gamma^r{}_{rr} &= \frac{R_{,rr} - R_{,r} \, \frac{E_{,r}}{E} -R \, \frac{E_{,rr}}{E} + R \left(\frac{E_{,r}}{E} \right)^2}{R_{,r} - R \, \frac{E_{,r}}{E}} -  \frac{1}{2}\frac{k_{,r}}{1-k} &
\Gamma^q{}_{qq} &= -\Gamma^p{}_{qq} = \Gamma^p{}_{pq} = \Gamma^p{}_{qp} = -\frac{E_{,q}}{E} 	\nonumber	\\
\Gamma^r{}_{pp} &= \Gamma^r{}_{qq} = -\frac{R}{E^2} \frac{1 - k}{R_{,r} - R \, \frac{E_{,r}}{E}}	&  
\Gamma^p{}_{pt} &= \Gamma^p{}_{tp} = \Gamma^q{}_{qt} = \Gamma^q{}_{tq} = \frac{R_{,t}}{R} \nonumber	\\
\Gamma^r{}_{rp} &= \Gamma^r{}_{pr} = \frac{E_{,r} E_{,p}/E^2  - E_{,rp}/E}{R_{,r}/R - E_{,r}/E}	&
\Gamma^p{}_{pr} &= \Gamma^p{}_{rp} = \Gamma^q{}_{qr} = \Gamma^q{}_{rq} = \frac{R_{,r}}{R} - \frac{E_{,r}}{E}	\nonumber	\\
\Gamma^r{}_{rq} &= \Gamma^r{}_{qr} = \frac{E_{,r} E_{,q}/E^2  - E_{,rq}/E}{R_{,r}/R - E_{,r}/E}  &  
&\:
\end{align}

The Riemann tensor can be summarized using the compactified functions defined in Eqs.~(\ref{eq:F}) and (\ref{eq:H}) by \cite{KoksbangRayTrace}
\begin{align}
\label{eq:Riemann}
\mathcal{R}_{trtr} &= -\frac{1}{2}H_{,tt} + \frac{H_{,t}^2}{4H} \nonumber	\\
\mathcal{R}_{tptp} = \mathcal{R}_{tqtq} &= -\frac{1}{2}F_{,tt} + \frac{F_{,t}^2}{4F} \nonumber	\\
\mathcal{R}_{rprp} &= -\frac{1}{2} (H_{,pp} + F_{,rr}) + \frac{1}{4}H_{,t}F_{,t} + \frac{1}{4H} (H_{,r}F_{,r} + H_{,p}^2) + \frac{1}{4F} (H_{,p}F_{,p} + F_{,r}^2 - H_{,q}F_{,q}) \nonumber	\\
\mathcal{R}_{rqrq} &= -\frac{1}{2} (H_{,qq} + F_{,rr}) + \frac{1}{4}H_{,t}F_{,t} + \frac{1}{4H} (H_{,r}F_{,r} + H_{,q}^2) + \frac{1}{4F} (-H_{,p}F_{,p} + F_{,r}^2 + H_{,q}F_{,q}) \nonumber	\\
\mathcal{R}_{pqpq} &= -\frac{1}{2} (F_{,pp} + F_{,qq}) + \frac{1}{4}F_{,t}^2 - \frac{1}{4H} F_{,r}^2 + \frac{1}{2F}(F_{,p}^2 + F_{,q}^2 )
\end{align}
\end{widetext}
The remaining nonzero terms can be found by the Riemann tensor's symmetry properties:
\begin{align}
\label{eq:Riemannsymmetry}
\mathcal{R}_{\alpha \beta \gamma \delta} &= -\mathcal{R}_{\alpha \beta \delta \gamma} = -\mathcal{R}_{\beta \alpha \gamma \delta} = \mathcal{R}_{\gamma \delta \alpha \beta}.
\end{align}

The Ricci scalar is simply
\begin{align}
\label{eq:Rscalar}
\mathcal{R} &= 8\pi \rho + 4 \Lambda
\end{align}

The Weyl curvature tensor is split into electric and magnetic parts as \cite{BolejkoSC,HellabyWormholes}
\begin{align}
\label{eq:Weyl}
E^\alpha{}_\beta &= \tensor{C}{^\alpha _\gamma _\beta _\delta} u^\gamma u^\delta \nonumber \\
 &= \frac{M \: R' - R \: M'/3}{R^3(R' - R \: E'/E)}\mathrm{diag}(0,2,-1,-1), \\
H_{\alpha \beta} &= \frac{1}{2} \epsilon_{\alpha \gamma \mu \nu} \tensor{C}{^\mu ^\nu _\beta _\delta}u^\gamma u^\delta = 0. 
\end{align}
The Kretschmann scalar, which is useful for identifying real singularities, is \cite{Walters}
\begin{align}
\label{eq:Kretschmann}
\mathcal{K} &= \mathcal{R}_{\alpha \beta \gamma \delta} \mathcal{R}^{\alpha \beta \gamma \delta} \nonumber \\
&= (8\pi)^2 \left( \frac{4}{3} \bar{\rho}_{\mathrm{int}}^2 - \frac{8}{3} \bar{\rho}_{\mathrm{int}} \rho +3 \rho^2 \right) + \frac{4}{3}\Lambda (2\Lambda + 8\pi \rho),
\end{align}
where $\bar{\rho}_{\mathrm{int}}$ is the ``average'' density inside shell $r$, defined in Eq.~(\ref{eq:densitycomparison}). This tells us that the spacetime singularities coincide with divergent densities, i.e. shell crossings and the bang/crunch.

The 3-spaces of constant time have non-zero Riemann tensor components \cite{HellabyPoS}
\begin{align}
\label{eq:3Riemann}
{}^3\mathcal{R}^r{}_{prp} &= {}^3\mathcal{R}^r{}_{qrq} = \frac{k'/2 - k \, \frac{E'}{E}}{R(R' - R \, \frac{E'}{E})}, \\
{}^3\mathcal{R}^p{}_{qpq} &= \frac{k}{R^2},
\end{align}
and a Ricci scalar \cite{BolejkoSCCMB}	
\begin{align}
\label{eq:3Rscalar}
{}^3\mathcal{R} &= 2\frac{k}{R^2}\left(\frac{ k'/k - 2 E'/E}{R'/R -  E'/E} + 1 \right),
\end{align}
Note that the model is perfectly spatially flat if $k = 0$, even if there is significant inhomogeneity.

\section{\label{app:rotationproof}Demonstration of shell rotation in metric}

We are interested in the geometry on a constant-$t$ hypersurface. We begin with the LT metric in spherical coordinates:
\begin{align}
\mathrm{d}s^2 &= -\mathrm{d}t^2 + \frac{(R')^2}{1-k} \mathrm{d}r^2 + R^2(\mathrm{d}\theta^2 + \sin^2 \theta \, \mathrm{d}\phi^2).
\end{align}
The shell shifting effect (section~\ref{sec:shifting}) modifies the separation between shells in the radial direction (orthogonal to the shell surface) by replacing $R'$ with $R' - R \: E'/E$. Because this change is in the radial direction, the actual distance is affected by the curvature, and we have built this into our definition of the shifting. The shift also adds a transverse separation between points on different shells, though, even when the angular coordinates are held constant. Because this component of the distance is along the shell's surface, the curvature function does not play a role. 

\begin{figure}[tbp]
\begin{center}
\includegraphics[width=8.5cm]{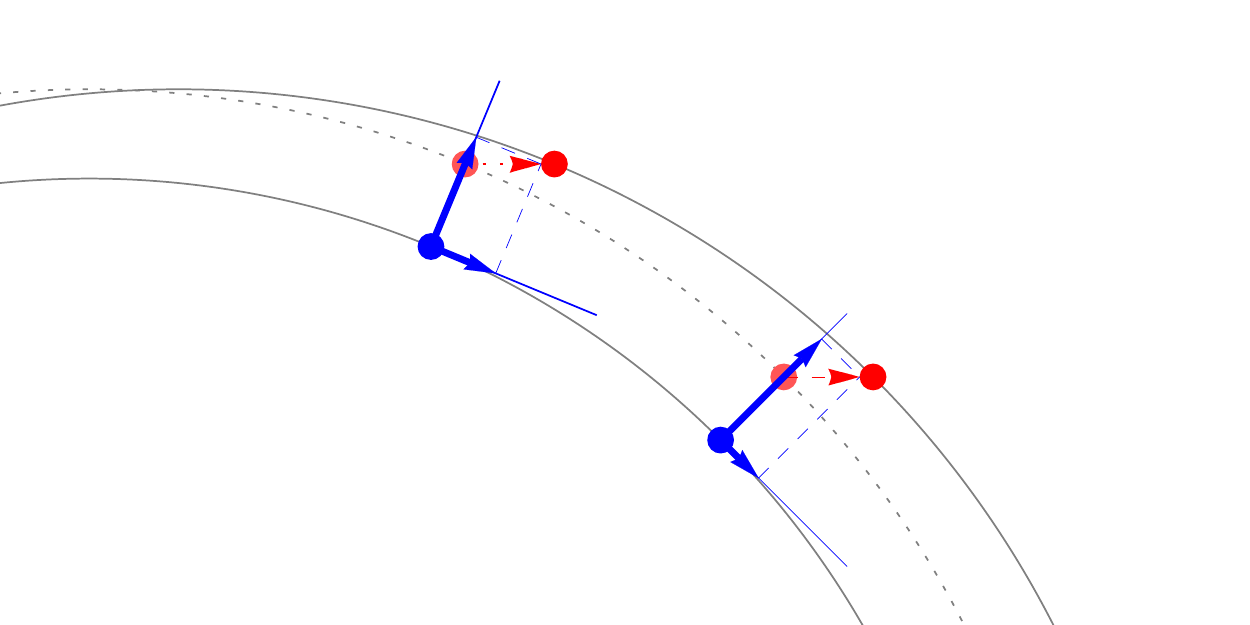}	
\caption[ShiftSeparation]{The separation between nearby points on different shells without accounting for shell rotation. The dotted arc shows where the outer shell would be without shifting; the red arrow shows how the individual points on the outer shell have moved. We have used $\mathrm{d}\theta = 0$ for illustrative purposes, so the red and blue dots in each set have the same $\theta$, measured from the top. The blue lines coming from the blue dots show the radial and transverse ($\theta$) directions, and the blue arrows and dashed lines show the components of the total separation in each direction. Two sets of points are shown to demonstrate the $\cos \theta$ factor of the $P'/S$ term in Eq.~(\ref{eq:shifttheta}) (in the limit $\delta r \rightarrow 0$).}
\label{fig:ShiftSeparation}
\end{center}
\end{figure}

Accounting only for the shifting effect, the points $(r, \theta, \phi)$ and $(r + \mathrm{d}r, \theta + \mathrm{d}\theta, \phi + \mathrm{d}\phi)$ have a total separation in the $\theta$ direction of
\begin{align}
\label{eq:shifttheta}
R \, \mathrm{d}\theta + R \left( \frac{P'}{S} \cos \theta \cos \phi + \frac{Q'}{S} \cos \theta \sin \phi - \frac{S'}{S} \sin \theta \right) \, \mathrm{d}r,
\end{align}
and in the $\phi$ direction of
\begin{align}
\label{eq:shiftphi}
R \sin \theta \, \mathrm{d}\phi + R \left( -\frac{P'}{S} \sin \phi + \frac{Q'}{S} \cos \phi \right) \, \mathrm{d}r.
\end{align}
A simple diagram of this decomposition is shown in Fig.~\ref{fig:ShiftSeparation}.

The total square separation is simply the sum of the squares of the components in these orthogonal directions (because space is approximately Euclidean on sufficiently small scales). This would give
\begin{widetext}
\begin{align}
\mathrm{d}s^2 &= -\mathrm{d}t^2 + \left[\frac{(R' - R\frac{E'}{E})^2}{1-k} + R^2 \left( \frac{P'}{S} \cos \theta \cos \phi + \frac{Q'}{S} \cos \theta \sin \phi - \frac{S'}{S} \sin \theta \right)^2 + R^2 \left( - \frac{P'}{S} \sin \phi + \frac{Q'}{S} \cos \phi \right)^2 \right] \mathrm{d}r^2 \nonumber \\
 &\ + 2 R^2 \left(\frac{P'}{S} \cos \theta \cos \phi + \frac{Q'}{S} \cos \theta \sin \phi - \frac{S'}{S} \sin \theta \right) \mathrm{d}r \, \mathrm{d}\theta + 2 R^2 \sin \theta \left( -\frac{P'}{S} \sin \phi + \frac{Q'}{S} \cos \phi \right) \mathrm{d}r \, \mathrm{d}\phi \nonumber \\
&\ + R^2(\mathrm{d}\theta^2 + \sin^2 \theta \, \mathrm{d}\phi^2).
\end{align}
But this does not match the metric as given in Eq.~(\ref{eq:sphericalmetric}). Shell shifting alone is not enough.

The rotation described in section~\ref{sec:rotation} adds a further transverse separation between the two points. This amounts to an additional
\begin{align}
\label{eq:rottransverse}
R \left( - \frac{P'}{S} \cos \phi - \frac{Q'}{S} \sin \phi \right) \, \mathrm{d}r, \qquad R \left( \frac{P'}{S} \cos \theta \sin \phi - \frac{Q'}{S} \cos \theta \cos \phi \right) \, \mathrm{d}r.
\end{align}
in the $\theta$ and $\phi$ directions, respectively. Incorporating both effects, the $rr$ part of the metric becomes
\begin{align}
g_{rr} &= \frac{(R' - R\frac{E'}{E})^2}{1-k} + R^2 \left[ \frac{P'}{S} (\cos \theta - 1) \cos \phi + \frac{Q'}{S} (\cos \theta - 1) \sin \phi - \frac{S'}{S} \sin \theta \right]^2 + R^2 (1 - \cos \theta)^2 \left( \frac{P'}{S} \sin \phi - \frac{Q'}{S} \cos \phi \right)^2 \nonumber \\
&= \frac{(R' - R\frac{E'}{E})^2}{1-k} + R^2 (1 - \cos \theta)^2 \left( \frac{P'^2 + Q'^2 + S'^2}{S^2} \right) + 2 R^2 (1 - \cos \theta) \left(\frac{P'S' \sin \theta \cos \phi + Q'S' \sin \theta \sin \phi + S'^2 \cos \theta}{S^2} \right).
\end{align}
\end{widetext}
With some simplification, we can confirm that this matches the corresponding term in Eq.~(\ref{eq:sphericalmetric}). Similarly, we can find $g_{r\theta}$ from Eqs.~(\ref{eq:shifttheta}) and (\ref{eq:rottransverse}):
\begin{align}
g_{r\theta} &= R^2 (\cos \theta - 1) \left( \frac{P'}{S} \cos \phi + \frac{Q'}{S}  \sin \phi \right) - R^2 \frac{S'}{S} \sin \theta \nonumber \\
&= R^2 \frac{1 - \cos \theta}{\sin \theta} \frac{E'}{E} + R^2 \left( \frac{\cos \theta - \cos^2 \theta}{\sin \theta} - \sin \theta \right)\frac{S'}{S}.
\end{align}
Again, this matches Eq.~(\ref{eq:sphericalmetric}). Finally, the $r\phi$ term:
\begin{align}
g_{r\phi} &= R^2 \sin \theta (1 - \cos \theta) \left( \frac{Q'}{S} \cos \phi - \frac{P'}{S} \sin \phi \right).
\end{align}

Thus, we see that the shell shifting and rotation effects fully encapsulate the differences between the LT and Szekeres metrics.

\section{\label{app:example}Example model definition}

Here we detail the function definitions used in the model shown in Figs.~\ref{fig:Szekeffects} and \ref{fig:expansionrates}.

We begin with a background FLRW model with $\linebreak H_0 = 70 \; \mathrm{km \; s^{-1} \; Mpc^{-1}}$, $\Omega_{\Lambda,0} = 0.7$, and $\Omega_{m,0} = 0.3$. The current age of the universe $t_0$ is determined by integrating the first Friedmann equation up to a scale factor of $1$: 
\begin{align}
t_0 = \int_0^1 \frac{\mathrm{d}a}{H_0 \sqrt{\Omega_{m,0} a^{-1} + \Omega_{\Lambda,0} a^2}},
\end{align}
which comes out to approximately $13.47$ Gyr.

\begin{figure}[tbp]
\begin{center}
(a)\\ \includegraphics[width=8.5cm]{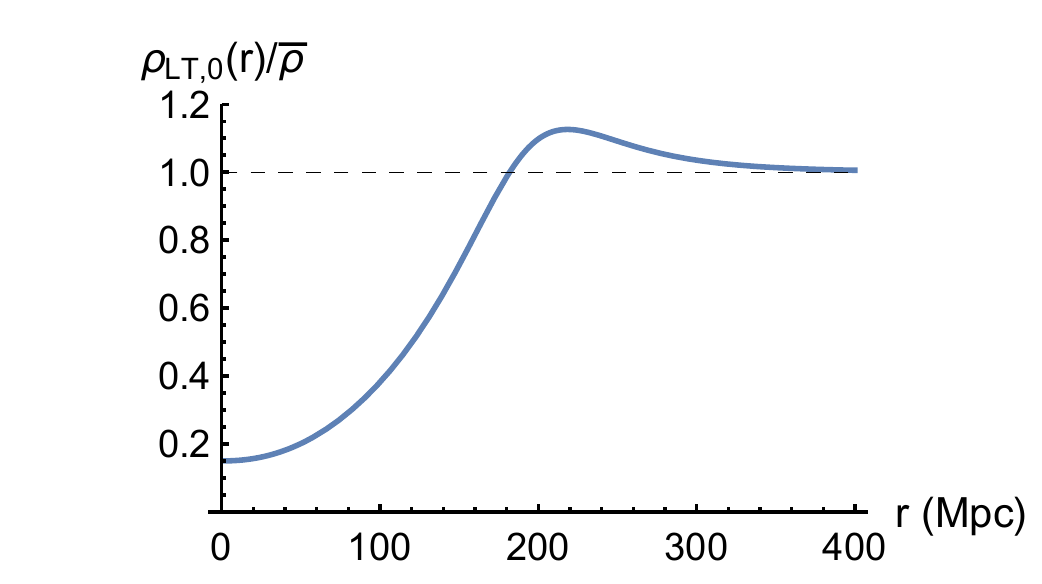}	\\%\hspace{0.4cm}
(b)\\ \includegraphics[width=8.5cm]{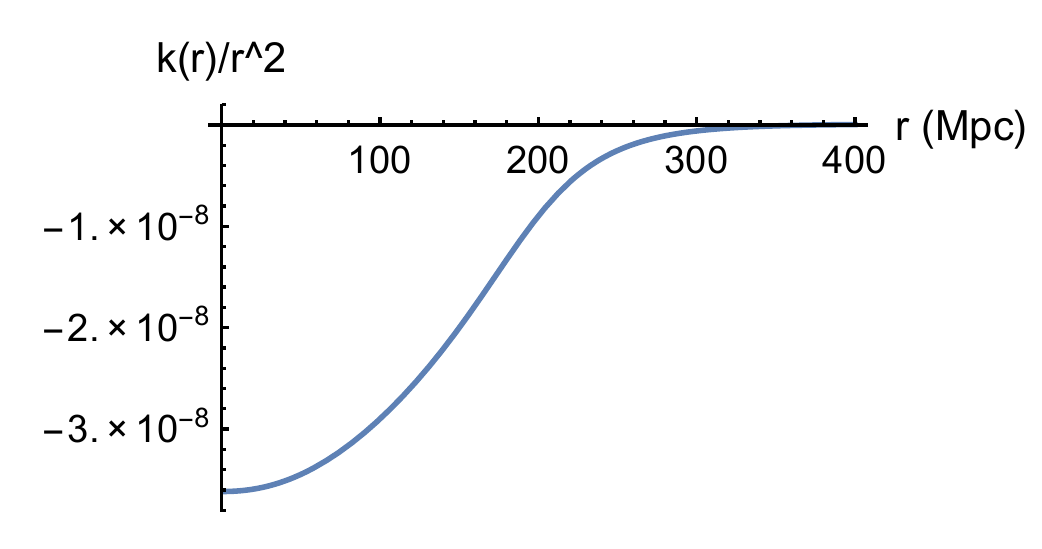}	\\%\hspace{0.4cm}
(c)\\ \includegraphics[width=8.5cm]{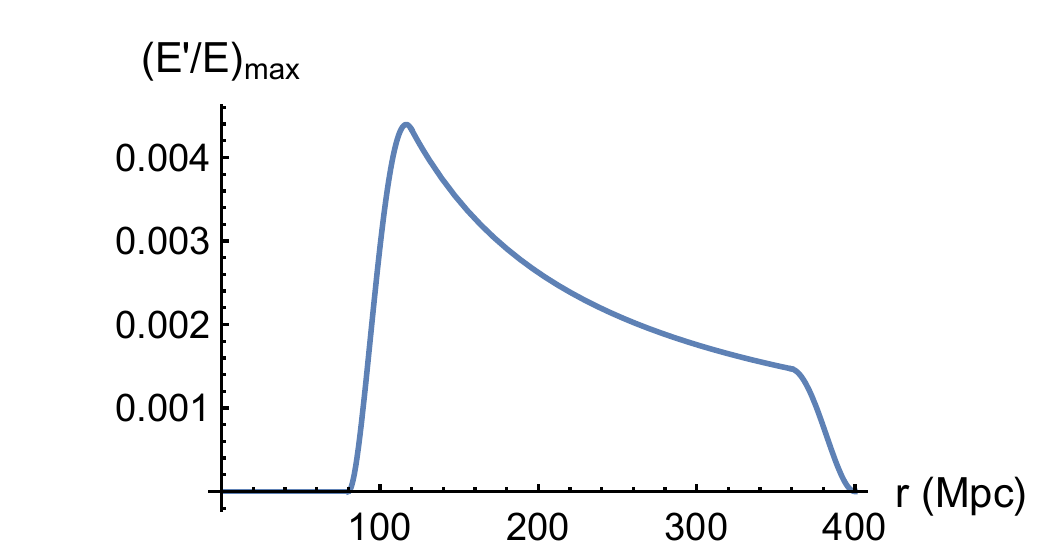}	
\caption[Example]{The density (a), curvature (b), and $(E'/E)_{max}$ (c) functions of the example models used in Figs.~\ref{fig:Szekeffects} and and \ref{fig:expansionrates}.}
\label{fig:example}
\end{center}
\end{figure}

We then define an LT model that matches onto this background at large $r$. We use a radial coordinate scaled so that $R(t_0,r) = r$ in units of Mpc, and set ${t_B(r) = 0}$. The density profile at $t_0$ is the universal void density profile found by Hamaus {\itshape et al.} \cite{Hamaus}:
\begin{align}
\rho(r) = \bar\rho \left(1 + \delta_c \frac{1-(r/r_s)^\alpha}{1+(r/r_v)^\beta} \right),
\end{align}
where $\bar\rho$ is the background FLRW density, $\delta_c$ is the central density contrast, $r_v$ characterizes the size of the void, $r_s$ gives a scale radius at which the density equals the background, and $\alpha$ and $\beta$ determine the inner and outer slopes of the void's wall, respectively. We have used $\delta_c = -0.85$, $r_v = 200$, $r_s = 182$, $\alpha = 2.18$, and $\beta = 9.482$. The density function is plotted in Fig.~\ref{fig:example}(a). From the density, we obtain $M(r)$ by integrating:
\begin{align}
M(r) &= 4 \pi \int_0^r{\rho(r) \, r^2 \mathrm{d}r}.
\end{align}
The curvature function $k(r)$ is then fixed by solving Eq.~(\ref{eq:bangtime}) at $t = t_0$. Because the other functions are already defined, at any given $r$ there is a unique value of $k(r)$ which satisfies Eq.~(\ref{eq:bangtime}), which can be found numerically. We do this for a series of $r$ values, in increments of $1$, and interpolate to obtain a smooth function, plotted in Fig.~\ref{fig:example}(b).

We then introduce anisotropy through the dipole functions. We define $(E'/E)_{m}$ as a piecewise function such that $r \, (E'/E)_{m}$ starts at $0$, ramps up smoothly starting at $r = 80$, plateaus, and then ramps back down to $0$ at $r = 400$.
\begin{align}
\label{eq:exampleEpE}
\left(\frac{E'}{E}\right)_{m} &= \frac{C_P}{(1+r)^{0.99}} \begin{cases}
		0 & r \leq 80, \\
		\left(1 - \frac{(r-120)^2}{1600} \right)^2 & 80 < r \leq 120, \\
		1 & 120 < r \leq 360, \\
		\left(1 - \frac{(r-360)^2}{1600} \right)^2 & 360 < r \leq 400, \\
		0 & r > 400.
	\end{cases}
\end{align}
This is plotted in Fig.~\ref{fig:example}(c). The $(1+r)^{-0.99}$ factor was taken from the model used in \cite{BolejkoCoarse}, and $C_P$ is an overall strength factor, where a value of 1 would push the shells very close to a shell crossing. In this case, we used ${C_P = 0.5}$. 

For Fig.~\ref{fig:Szekeffects}, we use only $P(r)$, keeping $Q(r)$ at $0$ and $S(r)$ at $1$. This means that $(E'/E)_{m} = P'(r)$. To obtain $P(r)$, we simply integrate. For Fig.~\ref{fig:expansionrates}, we instead only use $S(r)$, keeping $P(r)$ at 0, meaning $(E'/E)_{m} = S'/S$. To obtain $S(r)$, we use $(\ln S)' = S'/S$, integrate, and take the exponent.

The geodesic shown in Fig.~\ref{fig:Szekeffects} was generated backwards from $t_0$ with the initial values
\begin{subequations}
\begin{align}
r_0 &= 400, \quad p_0 = P(400) + \cot \frac{5\pi}{12}, \quad q_0 = 0, \\
k^\mu &= \left\{-1, \frac{\cos \frac{7\pi}8}{\sqrt{g_{rr}}}, -\frac{\sin \frac{7\pi}8}{\sqrt{g_{pp}}}, 0\right\}. 
\end{align}
\end{subequations}

The model used in the embedding illustration, Fig.~\ref{fig:embedding}, is similar, but with some changes to the parameters. The background FLRW model has a strong positive curvature, with ${\Omega_{m,0} = 2}$, and the void central density contrast is only $\delta_c = -0.3$, so that the curvature is positive even in the void. Also, $\Omega_{\Lambda,0} = 0$ and $H_0 = 300 \; \mathrm{km \; s^{-1} \; Mpc^{-1}}$. The dipole functions are axially symmetric, following Eq.~(\ref{eq:exampleEpE}) with $C_P = 0.7$.

%\clearpage

\end{document}